\documentclass[onecolumn,11pt,prd,nofootinbib]{revtex4-1}
\usepackage{amsmath,graphicx,amssymb,slashed,array}
\usepackage[math]{cellspace}
\begin{document}
\title{Polarization of a vector boson produced in decay of a heavy fermion in an 
arbitrary frame}
\author{Arunprasath V.} 
\email{arunprasath@imsc.res.in}
\affiliation{Institute of Mathematical Sciences,  CIT Campus, 
Taramani, Chennai, 600 113, India.}
\author{Ritesh K. Singh}
\email{ritesh.singh@iiserkol.ac.in}
\affiliation{Department of Physical Sciences, Indian Institute of Science
Education and Research Kolkata, Mohanpur, 741246, India}

\begin{abstract}
We consider processes where an electroweak gauge boson ($W$ and $Z$) is produced in the decays of a heavy fermion.
The polarization state of the produced gauge boson is given in terms of a vector polarization and a rank-2 tensor
polarization. In the rest frame of the mother fermion, these are given directly 
by the dynamical parameters of the underlying theory. In a frame where the mother particle is moving,
the polarization parameters of the gauge boson are dependent additionally upon kinematical factors.
We show that these kinematical factors depend only on the magnitude of the velocity of the mother particle
and derive analytical expressions for them. We apply the results to pair production of heavy fermions at the LHC
with one of the fermions decaying to a gauge boson ($W$, $Z$) and a light Standard Model fermion. We construct estimators of laboratory frame
values of polarization parameters of the produced gauge boson. These estimators can be used to estimate the laboratory frame values of 
polarization parameters of the produced gauge boson without a detailed simulation of the entire process. We validate our 
expressions with detailed Monte Carlo simulations in the context of beyond the Standard Model scenarios which have a vector-like top partner. We also indicate how to 
include finite width effects of the heavy fermion in some special cases.
\end{abstract}
\maketitle
\section{Introduction}\label{sec:1}

   Polarization of unstable particles such as 
the top, the tau or the electroweak gauge bosons ($W,Z$) produced in high energy colliders is a powerful
   tool. It can be used to probe the production or decay of these particles.  In particular,
   the top quark's polarization has been widely suggested as a probe of new 
physics~\cite{Kane:1991bg,Jezabek:1994zv,Godbole:2006tq,Baumgart:2013yra,Perelstein:2008zt,Arai:2010ci,Gopalakrishna:2010xm,Godbole:2010kr,Bhattacherjee:2012ir,Hikasa:1999wy,Godbole:2011vw,Krohn:2011tw,Rindani:2011pk,Choudhury:2010cd,Belanger:2013gha,Biswal:2012dr,Huitu:2010ad,Cao:2011hr,Fajfer:2012si,Rindani:2011gt}.  
There exist works on the polarization of tau lepton~\cite{Nojiri:1994it}
   and the gauge bosons ($W$, $Z$)\cite{AguilarSaavedra:2006fy,AguilarSaavedra:2010nx,Rahaman:2016pqj,Rahaman:2017aab} as well. The polarization of a particle depends not only on 
the physics involved in its production but also on the kinematics of its 
production~\cite{Belanger:2012tm,Shelton:2008nq,V.:2016wba}. In this work, we investigate the effect of kinematics on the polarization state of a spin-1 particle, such as $W$ or $Z$, 
in a class of processes where the boson is produced in the decays of a heavy fermion.

   Consider the production of a gauge boson $W$ at the LHC. It can be produced directly: e.g, $pp\rightarrow W^{+}W^{-}$. It can as well be produced in 
decays of
   a heavy particle: e.g., $pp\rightarrow t\bar{t}$ with $t\rightarrow bW$ and $\bar{t}$ decaying inclusively. The 
   kinematics of $W$ production in the second case is different from that in the first case. In the second case, there is a Lorentz boost from the rest frame of the top to the laboratory(lab) frame.
   This boost arises since the top quark, the parent particle of the $W$, moves with a non zero velocity in the lab frame. The effect of this boost on the value of $W$ polarization
   measured in the lab frame in non trivial. To see this, we first note that the 
   polarization of $W$ is determined by the underlying theory in the rest frame of the top.   
    Since the helicity
  states of $W$ are not invariant under arbitrary Lorentz boosts(see, for example \cite{Bourrely:1980mr,Boudjema:2009fz}),
   the polarization of $W$ measured in the lab frame is, in general, different from its value in the top quark rest frame. 
   
   In this work, we generalize the above mentioned example to include all processes of the type
 
    \begin{equation}\label{eq:1}
     pp\rightarrow F\bar{F}\rightarrow \bar{F}V f 
   \end{equation}
   where $F$ denotes a heavy quark, $V$ a vector boson ($W$, $Z$), and $f$ a light quark\footnote{The restriction to pair production of $F$ is dictated by the requirement that 
the fermion $F$ be unpolarized (see below). Since the pair production of these vector-like quarks is dominantly a QCD process, the produced vector-like quarks are unpolarized. This argument applies to the top quark pair production as well.}.. We assume that  the $\bar{F}$ decays inclusively. As will be discussed below,  the lab frame value of polarization 
of the particle  is related to that in the rest frame of its mother particle by a kinematical factor. This  opens up the possibility of predicting the lab frame polarization, given the theory.  In an earlier work~\cite{V.:2016wba} we considered
   the polarization of a spin-1/2 particle (the top quark) in cases where the particle is produced in the decay of a heavy spin-1/2 or a spin-0 particle. This work may be considered as an extension
   of our previous work to the case of a spin-1 boson (such as $W$ or $Z$) produced in the decay of a heavy quark.

     The gauge boson $V$ produced in the decays of the heavy quark $F$ 
   further decays to SM fermions according to its polarization state. Since the 
vector bosons are spin-1 particles, the polarization state is described by a 
three component vector ($P_{i}$ $(i=x,y,z)$) and a five component tensor 
polarization ($T_{xy},T_{yz},T_{zz},T_{xx}-T_{yy},T_{xz}$). In the rest frame of 
the heavy quark $F$,  the polarization parameters of $V$ are given by dynamical parameters such as 
couplings, mixing matrices and masses of the particles involved in the decay. On the other hand, 
 the lab frame values of polarization parameters of $V$ are simpler to be measured in experiments as they do not require the full reconstruction of the entire event. 
 
 To connect the experimentally  measured values of polarization parameters with theory parameters (the dynamical parameters relevant in the decay of $F$), one requires a method to predict 
 the lab frame values of the polarization parameters from a theory. To achieve this, we propose estimators for lab frame polarization parameters of $V$~\cite{V.:2016wba}.
 
   These estimators (for the eight polarization parameters of $V$) are of the 
following form:
\begin{equation}\label{eq:2}
\mathcal{P}_{estimator} \equiv \frac{1}{\sigma_{FF}}\int \frac{d\sigma_{FF}}
{d\beta_{F}} \ \ \mathcal{P}(\beta_{F}).
\end{equation} 
Here, $\sigma_{FF}$ is the cross section for the pair production of the parent 
quark $F$: $pp\rightarrow F\bar{F}$ and $\beta_F$ is the velocity of $F$ in the 
lab frame. $\mathcal{P}(\beta_{F})$ can be interpreted as the polarization 
parameter of $V$ in a single event.  Then the above expression can be 
interpreted as a weighted average of a polarization parameter of $V$ over the 
entire sample of events. The weighting factor $\frac{d\sigma_{FF}}{d\beta_{F}}$ 
includes convolution over the parton distribution functions.
Note that these estimators require only the magnitude of velocity of its mother particle. Hence, these estimators are simple to use in a Monte Carlo simulation of the relevant process. 
An advantage of the use of these estimators is that one can now predict the value of polarization parameters of $V$ without any simulation of the decay of $F$, its mother particle.  
These estimators are derived under the assumption that the parent quark has a narrow width ($\Gamma/m_F\ll 1$). We shall later relax this assumption and consider the case 
of larger widths of the mother particle in Sec.~\ref{sec:5}. We find that these estimators reproduce the polarization parameters of $V$ to within a few percent in the narrow width case. 
In the finite width case, when the non resonant and off-shell contributions are small, we modify these estimators to include the Breit-Wigner shape of the propagator of $F$. 
We find that both the original and the modified estimators provide equally good approximations to the polarization parameters of $V$, in the finite width case.

The processes mentioned above are possible in models with heavy fermions with electroweak gauge 
couplings. The Standard Model 
   top quark pair production and decay ($pp\rightarrow t\bar{t}\rightarrow \bar{t}bW$) also belongs to this class of processes. The heavy fermion $F$ may belong to a chiral or a vector-like
   representation of the electroweak gauge group. In this work, we take either $F$ to be a hypothetical vector-like quark or the top quark\footnote{The simplest model involving additional chiral heavy
   fermions, the fourth generation SM,  has been ruled out by the discovery of 
the Higgs at the LHC~\cite{Djouadi:2012ae}. The reason for the exclusion is the 
accidental suppression of $h\rightarrow\gamma\gamma$ rate by two orders of 
magnitude relative to the SM value, in this model. However, the possibility of the existence of fermions in vector-like 
representations of the SM gauge group is not excluded, due to the decoupling 
properties of vector-like 
fermions~\cite{Cacciapaglia:2010vn,Cacciapaglia:2011fx,Aguilar-Saavedra:2013qpa,Ellis:2014dza}.}. Examples of such processes involving vector-like fermions are: $pp\rightarrow T\bar{T}\rightarrow \bar{T}Zt$ and  $pp\rightarrow B\bar{B}\rightarrow \bar{B}tW$ where $T$ ($B$) is a vector-like quark with charge +2/3 (-1/3). The formalism used in this work
is applicable to any model of heavy fermions with such decay processes.

   Vector-like fermions appear naturally in various beyond the SM scenarios 
which address the hierarchy problem in the Higgs sector: warped extra-dimension 
models~\cite{Carena:2006bn}, composite Higgs 
models~\cite{Agashe:2004rs,Contino:2006qr,Contino:2006nn} and Little Higgs 
models~\cite{ArkaniHamed:2001nc,ArkaniHamed:2002pa,ArkaniHamed:2002qx,
ArkaniHamed:2002qy,Low:2002ws,Perelstein:2003wd}, for example. There is a renewed interest in the phenomenology of 
vector-like fermions. Recently, a number of studies that were proposed to 
explain a possible evidence of a 750 GeV resonance decaying to 
diphoton final state~\cite{TheATLAScollaboration:2015mdt,CMS:2015dxe} considered models with additional vector-like 
fermions (see, for example~\cite{DiChiara:2015vdm,Franceschini:2015kwy,Harigaya:2015ezk,Buttazzo:2015txu,Ellis:2015oso}).

The vector-like 
fermions can have a bare mass term unlike the chiral fermions of the SM. As a 
result, one can have heavy vector-like fermions without a large coupling to 
Higgs(if allowed by symmetry). This leads to the decoupling of their effects in 
electroweak oblique corrections, Higgs production and decay, in the limit where 
their mass goes to infinity, with their coupling to Higgs (if any) remains 
fixed. Perturbative unitarity bounds can be also evaded for any given large mass 
of a vector-like fermion, for an appropriately small mixing with the SM 
fermions~\cite{Dawson:2012di}. 

   Though vector-like fermions can appear in various representations of the SM 
gauge group, the strongly interacting fermions of charge +2/3 and -1/3, the 
so-called top quark and bottom quark partners play important role in the models mentioned 
above.  For example, in Little Higgs models, the quadratic divergence in Higgs 
self-energy coming from the top quark loop is canceled by the contribution of 
top partners.

The masses of these vector-like quarks are constrained by the direct searches at 
the LHC~\cite{Sirunyan:2017pks,Sirunyan:2017usq}. The assumption is that they 
couple only to third generation SM quarks. In general, vector-like quarks can couple to
the first two generations of SM 
fermions as well~\cite{Cacciapaglia:2011fx}. Depending upon the assumptions on 
the branching ratios, the lowest direct search bounds from the LHC read 790 GeV 
for the charge +2/3 quark, 730 GeV for the charge -1/3 
quark~\cite{Sirunyan:2017usq}. 

   Flavor observables, precision electroweak observables, Higgs coupling 
measurements constrain indirectly the mixing  of the top quark (or bottom quark ) partners with the SM quarks 
 and their masses. The constraints depend upon the representation of 
these vector-like quarks under the SM gauge 
group~\cite{Cacciapaglia:2010vn,Cacciapaglia:2011fx,Aguilar-Saavedra:2013qpa,
Ellis:2014dza}.

   This paper is divided into six sections with the first section being the introduction to this work. 
   Section~\ref{sec:4} describes the expressions of polarization parameters of $V$ in the rest frame of $F$. 
   Section~\ref{sec:3} discusses two models of a vector-like quark $T$ of charge +2/3. In this section, benchmark values of model parameters are provided  and the corresponding $Z$ polarization 
   parameters are obtained for the decay $T\rightarrow Z t$.   
  
   Section~\ref{sec:2} describes the formalism for the derivation of the polarization 
   estimators. Section~\ref{sec:5}
   completes the derivation of polarization estimators of $V$ at the level of $pp$ collisions and describes numerical validation of the polarization estimators. Section~\ref{sec:6} presents 
   a summary of this work.

\section{Polarization parameters of V}\label{sec:4}
 In this section, we give expressions for the polarization parameters of the 
vector boson $V$, in the rest frame of the parent particle $F$ in the decay 
$F\rightarrow Vf$. The only non-vanishing
polarization parameters are $P_z$ and $T_{zz}$ and they are defined by:
\begin{align}\label{eq:40}    
P_z&=\frac{\Gamma^{+}-\Gamma^{-}}{\Gamma^{+}+\Gamma^{-}+\Gamma^{0}},\\\nonumber    
T_{zz}&=\frac{\Gamma^{+}-2\Gamma^{0}+\Gamma^{-}}{\Gamma^{+}+\Gamma^{-}+\Gamma^{0
}}
  \end{align}
  where $\Gamma^{i}$ ($i=+,0,-$) denote the partial decay widths corresponding 
to the decay of $F$ into an unpolarized $f$ and an electroweak gauge boson $V$ 
with helicity $i$. Taking
  the vertex $FfV$ vertex as $\gamma^{\mu}(g_L P_L + g_R P_R)$ where $P_L$ and 
$P_R$ are the left and right chiral projectors, we get,
\begin{widetext}
  \begin{align}\label{eq:41}  
P_z&=\frac{2(g_R^2-g_L^2)K^{1/2}\xi_V}{-12\xi_V\sqrt{\xi_f}
g_Lg_R+(g_L^2+g_R^2)(K+3\xi_V(1+\xi_f-\xi_V))},\\\nonumber
T_{zz}&=-\sqrt{\frac{2}{3}}\frac{(g_R^2+g_L^2)K}{(-12\xi_V\sqrt{\xi_f}
g_Lg_R+(g_L^2+g_R^2)(K+3\xi_V(1+\xi_f-\xi_V)))}    
  \end{align}
 \end{widetext}
where $\xi_V=m_V^2/m_F^2$, $\xi_f=m_f^2/m_F^2$ and $K\equiv K(1,\xi_V,\xi_f)$ 
with $K(x,y,z)=x^2+y^2+z^2-2xy-2yz-2zx$. $m_F$, $m_V$ and $m_f$ denote the 
masses of the particles $F$, $V$ and $f$, respectively.
Note that $T_{zz}$ is non zero even when $g_L=g_R$. 

\section{Models}\label{sec:3}  
     In this section we consider the models which have a vector-like quark ($T$) 
of charge +2/3 mixing with the third generation quarks of the 
SM~\cite{Cacciapaglia:2010vn,Aguilar-Saavedra:2013qpa}. The models depend upon 
the representation
 of $T$ under the SM gauge group. We consider the cases where $T$ is an 
electroweak singlet, $T$ forms an electroweak doublet with a vector-like quark 
$B$ of charge -1/13. We refer to
 them as the Singlet Model and the Doublet Model, respectively. In both the 
models, we consider the decay: $T\rightarrow Zt$ where $t$ is the top quark.
\subsection{The Singlet Model}\label{ssec:3a} 
 The Yukawa part of the Lagrangian reads
\begin{equation}\label{eq:36}
 \mathcal{L} = -y_t\bar{q}_L'\tilde{\Phi}t_R'-\lambda\bar{q}_L'\tilde{\Phi}T_R'- 
M\bar{T}_L'T_R'+h.c.
\end{equation}
 The Yukawa coupling of the top is denoted as $y_t$. We have not included the 
Yukawa part of the SM in the above equation. The primes refer to the fact that 
the Lagrangian is in gauge basis. $\Phi$ is the Higgs doublet with 
$\tilde{\Phi}=i\tau_2\Phi^{\ast}$ ($\tau_2$ is a Pauli matrix, $\ast$ denotes 
complex conjugation). After the Higgs doublet 
 acquires a vacuum expectation value (vev), $\langle \Phi \rangle = 
(0,v/\sqrt{2})'$ (the prime denotes transpose), $T$ mixes with the SM fermions 
with the amount of mixing determined by $\lambda$. Neglecting the top mixing 
with the first two generation fermions,
 we get two mass eigen states ($t$, $T$) by performing a bi-diagonalization of 
the mass matrix. Taking the mass eigen values as the physical masses of the top 
and $T$, $m_t$ and $m_T$, respectively,
 we have
 \begin{align}\label{eq:37a}
  \frac{y_t^2v^2}{2}&=m_t^2\left(1+\frac{x^2}{M^2-m_t^2}\right), m_T^2= 
M^2\left(1+\frac{x^2}{M^2-m_t^2}\right),\\\nonumber
  \sin\theta_L&=\frac{Mx}{\sqrt{(M^2-m_t^2)^2+M^2x^2}},\, 
\sin\theta_R=\frac{m_t}{M}\sin\theta_L
 \end{align}
where $x=\lambda v/\sqrt{2}$. $\theta_L$ and $\theta_R$ are the mixing angles of 
the left and right chiral parts of the top and $T$, respectively, obtained after 
the bi-diagonalization
of the top-$T$ mass matrix. We take the independent parameters as $m_t,m_T$ and 
$\theta_L$. Constraints on this model coming from  contribution
to oblique parameters $S,T$ and $U$, loop level corrections to $Zb\bar{b}$ 
vertex, in the form of upper bound on the mixing read $\sin\theta_L<0.15$, for 
$m_T>750$ GeV at $95\%$ Confidence Level (C.L)~\cite{Aguilar-Saavedra:2013qpa}.

\subsection{The Doublet Model}\label{ssec:3b}
 The Yukawa part of the Lagrangian reads,
 \begin{align}\label{eq:38a}  
\mathcal{L}=&-y_t\bar{q}_L'\tilde{\Phi}t'_R-\lambda_t\bar{Q}_L'\tilde{\Phi}
t_R'-\lambda_d\bar{Q}_L'\Phi b_R'\\\nonumber
 &-y_b\bar{q}_L'\Phi b_R'-M\bar{Q}_LQ_R' + h.c.
 \end{align}
where $Q=(T,B)$ is a doublet of vector-like quarks and $y_b$ denotes the Yukawa 
coupling of bottom quark. $\lambda_t$ and $\lambda_b$ determine the $t$- $T$ and 
$b$-$B$ mixing,
respectively. We have neglected the mixing of third generation quarks with the 
corresponding quarks of the first two generations as we are not interested in 
the effects on flavor observables.
After the Higgs gets a vev, the top and $T$ mix, and $b$ and $B$ mix as well. 
With $x=\lambda_t v/\sqrt{2}$ and $x_b=\lambda_b v/\sqrt{2}$, we have,
\begin{align}\label{eq:39a}
&\sin\theta_R^u=\frac{Mx_t}{\sqrt{(M^2-m_t^2)^2+M^2x_t^2}},\, 
\sin\theta_L^u=\frac{m_t}{M}\sin\theta_R^u,\\\nonumber
&\frac{y_t^2v^2}{2}=m_t^2\left(1+\frac{x_t^2}{M^2-m_t^2}\right),
m_T^2=M^2\left(1+\frac{x_t^2}{M^2-m_t^2}\right),\\\nonumber
&\sin\theta_R^d=\frac{Mx_b}{\sqrt{(M^2-m_b^2)^2+M^2x_b^2}},\, 
\sin\theta_L^d=\frac{m_b}{M}\sin\theta_R^d \\\nonumber
&\frac{y_b^2v^2}{2}=m_b^2\left(1+\frac{x_b^2}{M^2-m_b^2}\right),
m_B^2=M^2\left(1+\frac{x_b^2}{M^2-m_b^2}\right), 
\end{align}
where $\theta_{L,R}^{u}$ and $\theta_{L,R}^{d}$ denote mixing angles of the left 
chiral and right chiral parts of the $t-T$ and $b-B$ pairs, respectively and 
$m_b$ and $m_B$ denote the
masses of the bottom quark and the $B$ quark, respectively. Constraints from the 
appearance of tree level corrections to the $Zb\bar{b}$ vertex, contribution to 
oblique parameters
and loop level contributions to $Zb\bar{b}$ vertex read $\sin\theta_R^{d}<0.06$ 
and $\sin\theta_R^u<0.13$ for $m_T>750$ GeV at 95\% 
C.L~\cite{Aguilar-Saavedra:2013qpa}.  The mass splitting between $B$ and $T$ is
also severely constrained to lie within a few 
GeVs~\cite{Aguilar-Saavedra:2013qpa}. We assume that $B$ is degenerate in mass 
with $T$: $m_B=m_T$. In this model, there are three  free parameters:
$m_T$, $\theta_R^d$ and $\theta_R^u$.

\subsection{Benchmark points}\label{ssec:3c}
 In this subsection, we describe our choices of parameter values that are used 
throughout this work. In the case of the Singlet Model and the Doublet Model, we 
take the 
 lower bound of the mass of $T$ as 900 GeV, based on the current direct search 
constraints. We take  $\theta_L=0.1$ in the Singlet Model due to the constraints 
mentioned in the previous section
 We set $m_B=m_T$, $\theta_R^u=0.09$ and $\theta_R^d=0.05$ in the Doublet 
Model, 
 to be consistent with the constraints. The values are tabulated in 
Table~\ref{tab:1}.
\begin{table}
\begin{ruledtabular}
\begin{tabular}{ccc}
parameter  & Singlet & Doublet  \\ \hline
$m_T$ (GeV)   & $\geq 900$ & $\geq 900$ \\
$m_B$ (GeV)  & $= m_T$    & $= m_T$\\
mixing angle(s)  &  $\theta_L=0.1$ & $\theta_R^u=0.09$\\
  &                 & $\theta_R^d=0.05$  
\end{tabular}
\end{ruledtabular}
\caption{List of parameters and their values in the two models, the Singlet and 
the Doublet models. }
\label{tab:1}
\end{table}
\subsection{Discussion}\label{ssec:3d}
We now discuss the value of $P_z$ and $T_{zz}$ for specific models described 
above for the decay $T\rightarrow Zt$, and $t\rightarrow bW$ in the SM.
In the case of Singlet Model, the coupling $ZtT$ which is responsible for the 
decay $T\rightarrow Zt$, is purely left chiral: 
$g_L=(g/(2\cos\theta_W))\sin\theta_L\cos\theta_L$ and $g_R=0$ 
where $\theta_W$ denotes the weak mixing angle and $g$ the $SU(2)_L$ gauge 
coupling. In the case of Doublet Model, the $ZtT$ vertex is 
purely right chiral: $g_L=0$ , 
$g_R=-(g/(2\cos\theta_W))\cos\theta_R^u\sin\theta_R^u$. For these two models, 
the expressions for $P_z$ and $T_{zz}$ can be obtained by the following 
replacements: $\xi_V\rightarrow \xi_Z=m_Z^2/m_T^2$, $\xi_f\rightarrow 
\xi_t=m_t^2/m_T^2$ and substituting the values of $g_L$ and $g_R$ in 
Eq.~\ref{eq:41}. 

In the case of the SM top decay, the coupling is purely left chiral: 
$g_L=g/\sqrt{2}$ and $g_R=0$. The expression for $P_z$ and $T_{zz}$ of $W$ can 
be obtained by the following
replacements $\xi_V\rightarrow \xi_W=m_W^2/m_t^2$,
$\xi_f\rightarrow \xi_b=m_b^2/m_t^2$
and substituting the values of $g_L$ and $g_R$ in 
Eq.~\ref{eq:41}.

\section{Formalism}\label{sec:2}    
    To obtain the expressions for the polarization estimators, we begin by 
looking at the parton level process of the form:
    \begin{equation}\label{eq:3}
     p_1p_2\rightarrow F\bar{F}\rightarrow \bar{F}Vf
    \end{equation}
where $p_1,p_2$ denote partons. We shall assume that the produced $V$ further decays to leptons: 
$V\rightarrow\ell\bar{\ell}'$. The parton level amplitude can be written as:  
  \begin{align}\label{eq:4}    
i\mathcal{M}&=\frac{1}{(p^2_F-m_F^2)+im_F\Gamma_F}\frac{1}{
(p^2_V-m_V^2)+im_V\Gamma_V}\\\nonumber
    &\times  i\mathcal{M}^{\prime}(V\rightarrow \ell\bar{\ell}^{\prime})_{\mu} 
(-iP^{\mu\nu})\\\nonumber
    &\times  i\mathcal{M}^{\prime}(F\rightarrow 
fV)_{\nu\alpha}(iP_{\alpha\beta})\\\nonumber
    &\times i\mathcal{M}^{\prime}(q\bar{q}/gg\rightarrow F\bar{F})_{\beta}
  \end{align}
  where the primes on amplitudes denote that the wave functions of the particles 
which are exchanged have not been included. The quantities $P^{\mu\nu}$ and 
$P_{\alpha\beta}$ denote 
  the numerators of the propagators of $V$ and $F$, respectively. The indices 
$\alpha$ and $\beta$ ($\alpha,\beta=1,\cdots 4$) denote the components of the 
spinors and Dirac matrices. 
  For clarity, the denominators of the propagators have been explicitly taken 
out of the expressions for the amplitudes. We have 
   \begin{align}\label{eq:5}    
P^{\mu\nu}&=-\sum_{i}\hat{\epsilon}(p_V^{PCM})^{(i)}\hat{\epsilon}(p_V^{PCM})^{
(i)\ast},\;\;i=-1,0,1,\\\nonumber      
P_{\alpha\beta}&=\sum_{\lambda}u(p_F^{PCM},\lambda)_{\alpha}\bar{u}(p_F^{PCM},
\lambda)_{\beta},\;\;\lambda=-1/2,+1/2  
  \end{align}
  where $i$, $\lambda$ and  $p_F^{PCM}$, $p_V^{PCM}$ are the helicities and 
momenta of the particles $F$ and $V$, respectively, measured in the parton 
center of mass (PCM) frame. 
  Substituting  these relations in Eq.~\ref{eq:4}, and defining 
  \begin{align}\label{eq:6}    
i\mathcal{M}_{F\bar{F}}(p_F^{PCM},p_{\bar{F}}^{PCM})_{\lambda}&=\sum_{\beta}\bar
{u}(p_F^{PCM},\lambda)_{\beta}i\mathcal{M}^{\prime}(q\bar{q}/gg\rightarrow F\bar{F})_{\beta},\\\nonumber
i\mathcal{M}_{fFV}(p_F^{PCM},p_f^{PCM},p_V^{PCM})_{i\lambda}&=\sum_{\alpha}
i\mathcal{M}^{\prime}(F\rightarrow 
fV)_{\nu\alpha}\hat{\epsilon}^{\nu(i)\ast}(p_V^{PCM})u(p_F^{PCM},\lambda)_{
\alpha},\\\nonumber   
i\mathcal{M}_{V\ell\bar{\ell}^{\prime}}(p_V^{PCM},p_{\ell}^{PCM},p_{\bar{\ell}^{
\prime}}^{PCM})^{(i)}&= \sum_{\mu}i\mathcal{M}^{\prime}(V\rightarrow 
\ell\bar{\ell}^{\prime})_{\mu}\hat{\epsilon}^{\mu(i)}(p_V^{PCM}),
  \end{align}
  we write the amplitude as
  \begin{equation}\label{eq:7}
    i\mathcal{M}\sim 
i\mathcal{M}_{F\bar{F}}(p_F^{PCM},p_{\bar{F}}^{PCM})_{\lambda}
i\mathcal{M}_{fFV}
(p_F^{PCM},p_f^{PCM},p_V^{PCM})^{(i)}_{\lambda}
i\mathcal{M}_{V\ell\bar{\ell}^{
\prime}}(p_{V}^{PCM},p_{\ell}^{PCM},p_{\bar{\ell}^{\prime}}^{PCM})^{(i)}.
  \end{equation}  
 The phase space element of the parton level process can be written in terms of 
2-body phase space elements as  
\begin{equation}\label{eq:8}  
d\Phi=\frac{dp^2_V}{2\pi}\frac{dp_F^2}{2\pi}d\Phi_{F\bar{F}}d\Phi_{fFV}d\Phi_{
V\ell\bar{\ell}^{\prime}}
\end{equation}
with
\begin{align}\label{eq:9}  
d\Phi_{F\bar{F}}&=(2\pi)^4\delta^{(4)}(P_{in}-p_{F}-p_{\bar{F}})\frac{d^3p_F}{
(2\pi)^32E_F}\frac{d^3p_{\bar{F}}}{(2\pi)^32E_{\bar{F}}},\\\nonumber  
d\Phi_{fFV}&=(2\pi)^4\delta^{(4)}(P_{F}-p_{f}-p_{V})\frac{d^3p_f}{(2\pi)^32E_f}
\frac{d^3p_{V}}{(2\pi)^32E_{V}},\\\nonumber   
d\Phi_{V\ell\bar{\ell}^{\prime}}&=(2\pi)^4\delta^{(4)}(P_{V}-p_{\ell}-p_{\bar{
\ell}^{\prime}})\frac{d^3p_{\ell}}{(2\pi)^32E_{\ell}}\frac{d^3p_{\bar{\ell}^{
\prime}}}{(2\pi)^32E_{\bar{\ell}^{\prime}}},\\\nonumber
\end{align}
 where $P_{in}$ denotes the sum of initial state parton momenta and $p^2_F, 
p^2_{V}$ denote the invariant masses of $F$ and $V$, respectively 
($p^2_F>0,p^2_V>0$). 
 Squaring the amplitude $i\mathcal{M}$, multiplying with the phase space element 
and flux factor, and averaging over initial state spin and color indices, 
 summing over intermediate state spin and color indices, we get the parton level 
cross section for the process as
 \begin{align}\label{eq:10}
  \hat{\sigma}=\frac{f_{avg}^{in}}{2E_12E_2|\vec{v}_1-\vec{v}_2|}\int 
\Delta_{BW}(p_V^2)\Delta_{BW}(p_F^2)\sum |\mathcal{M}|^2d\Phi 
\end{align}
where $f_{avg}^{in}$ denotes the spin and color averaging factor for the initial 
state and $E_{i},\vec{v}_{i}$ $(i=1,2)$ denote the energies and velocities of 
the initial state partons. 
In the above equation, $\Delta_{BW}(p^2)=((p^2-m^2)+m^2\Gamma^2)^{-1}$ denote 
the Breit-Wigner factors arising from the propagators. 
When the width of an intermediate particle is much smaller than its width, 
$\Delta_{BW}(p^2)$ can be replaced by a Dirac delta function as follows\footnote{This approximation 
breaks down when the mass difference between the mother particle and any one of the daughter particles is of the order of the width of the 
mother particle~\cite{Berdine:2007uv,Kauer:2007zc,Uhlemann:2008pm}.}:
\begin{equation}\label{eq:11}
  \Delta_{BW}(p^2)\rightarrow \frac{\pi}{m\Gamma}\delta(p^2-m^2).
\end{equation}
Assuming that this approximation, the so-called Narrow-Width Approximation 
(NWA), holds for both $F$ and $V$, using the factorized form of the phase space 
element, 
we write Eq.~\ref{eq:10}  as
\begin{equation}\label{eq:12}
  \hat{\sigma}=f_{avg}^{V}f_{avg}^{F}\int 
\sum_{\{\lambda,i\}}\hat{\sigma}_{\lambda^{\prime}\lambda}^{F\bar{F}}\frac{
d\Gamma^{FfV}_{\lambda\lambda';i'i}}{\Gamma_F}\frac{d\Gamma^{V\ell\bar{\ell}'}_{
ii'}}{\Gamma_V}
\end{equation}
where $\int 
\sum_{\lambda}d{\sigma}^{F\bar{F}}_{\lambda\lambda}=\hat{\sigma}^{F\bar{F}}$, 
the cross section for the pair production of $F$,
$\int \sum_{(\lambda,i)}d\Gamma_{\lambda\lambda;ii}^{FfV}$ is the partial decay 
width for the decay $F\rightarrow fV$,  $\int 
\sum_{i}d\Gamma_{ii}^{V\ell\bar{\ell}'}$ is the 
partial decay width for the decay $V\rightarrow\ell\bar{\ell}'$ and 
$f_{avg}^{V}$ and $f_{avg}^{F}$ denote the spin averaging factors included in 
the 
definitions of $d\Gamma^{V}$ and $d\Gamma^{F}$.

The Lorentz invariance of the phase space element factors allows evaluation of 
different parts of the squared amplitude in different frames. 
The pair production of $F$ can be evaluated in the PCM frame, the decay of $F$ 
in the rest frame of $F$ and the decay of $V$ in the rest frame of $V$. 
The rest frame of $V$ can be reached from the PCM frame through the 
transformation: 
\begin{equation}
h^{-1}(p_V^{PCM})\equiv 
\Lambda_z^{-1}(\beta_V^{PCM})R^{-1}(\theta_V^{PCM},\phi_V^{PCM})
\end{equation}
where $\beta_V^{PCM}$, $\theta_V^{PCM}$ and $\phi_V^{PCM}$ define the velocity 
and the direction of motion of $V$ in the PCM frame:
\begin{equation}
  \mathrm{PCM}\; 
\mathrm{frame}\xrightarrow{h^{-1}(p_V^{PCM})}V\;\mathrm{rest}\;\mathrm{frame}.
\end{equation}

The rest frame of $F$ can be reached from the PCM frame by a Lorentz 
transformation 
\begin{equation}
 h^{-1}(p_F^{PCM})\equiv \Lambda_z^{-1}(\bar{\beta}) 
R^{-1}(\bar{\theta},\bar{\phi}) 
\end{equation}

where $\bar{\beta},\bar{\theta},\bar{\phi}$ define the velocity and the 
direction of motion of $F$ in the PCM frame:
\begin{equation}
  \mathrm{PCM}\; 
\mathrm{frame}\xrightarrow{h^{-1}(p_F^{PCM})}F\;\mathrm{rest}\;\mathrm{frame}.
\end{equation}
This transformation transforms the momenta to the rest frame of $F$ : $p_V^{PCM}\rightarrow 
p_V^{F}$, $p_f^{PCM}\rightarrow p_f^{F}$. 
Under this transformation, the helicities of $F$ are unchanged as the Lorentz 
transformation is along its direction of motion. 
On the other hand, the helicity states of $V$ transform in the following way:
\begin{equation}\label{eq:13}
  \hat{\epsilon}^{(i)}(p_V^{PCM})\xrightarrow{h^{-1}(p_F^{PCM})} 
R_{ki}(\mathcal{R})\hat{\epsilon}^{(k)}(p_V^{F})\;\;(i,k=+,0,-)
\end{equation}
where $R$ is a rotation matrix corresponding to the rotation
\begin{equation}\label{eq:14}
  \mathcal{R}=h^{-1}(p_V^{F})h^{-1}(p_F^{PCM})h(p_V^{PCM}).
\end{equation}
In the above expressions, $p_V^{F}=\Lambda_z^{-1}(\bar{\beta}) 
R^{-1}(\bar{\theta},\bar{\phi})p_V^{PCM}$ is the momentum of $V$ in the rest 
frame of $F$, and $p_V^{PCM}$ and $p_F^{PCM}$ 
denote the PCM frame momenta of $F$ and $V$, respectively\footnote{To see that 
the transformation in Eq.~\ref{eq:12} defines a rotation, consider the 
application of the 
transformations on the momentum of $V$ in its rest frame: 
$p_V^{V}\xrightarrow{h(p_V^{PCM})}p_V^{PCM}\xrightarrow{h^{-1}(p_F^{PCM})}p_V^{F
}\xrightarrow{h^{-1}(p_V^{F})}p_V^{V}$ ($p_V^V=(m_V,\vec{0})$). 
The transformation takes the rest frame of $V$ to itself. Hence, it is a 
rotation on the rest frame of $V$. 
This can also be checked explicitly using the expressions for the Lorentz 
transformations.}.
 The expression for the matrix element after the transformations have been 
applied becomes,
\begin{equation}\label{eq:15}
  i\mathcal{M}= 
\sum_{\lambda,k,i} i\mathcal{M}_{F\bar{F}}(p_F^{PCM},p_{\bar{F}}^{PCM})_{\lambda}
R_{ki}
i\mathcal{M}_{fFV}(p_F^{F},p_f^{F},p_V^{F})^{(k)}_{\lambda}i\mathcal{M}_{
V\ell\bar{\ell}^{\prime}}(p_{V}^{V},p_{\ell}^{V},p_{\bar{\ell}^{\prime}}^{V})^{
(i)}.
\end{equation}
This implies that the expression for the parton-level cross section for the 
process becomes,
\begin{equation}\label{eq:16}
  \hat{\sigma}=f_{avg}^{F}f_{avg}^{V}\times
  \int 
\sum_{\{\lambda\},\{i\}}  d\sigma^{F\bar{F}}_{\lambda'\lambda}
R^{\dagger}_{i'k'}
\left(\frac{d\Gamma^{FfV}_{\lambda\lambda';k'k}}{\Gamma_F}\right)_{F}R_{ki}
\left(\frac{d\Gamma^{V\ell\bar{\ell}'}_{ii'}}{\Gamma_V}\right)_V
\end{equation}
where $\{\lambda\}$ and $\{i\}$ denote the set of helicity indices corresponding 
to  $F$ and $V$ that appear in the above expression. In the above expression, 
the subscripts $F$ and $V$ 
on $d\Gamma^{fFV}/\Gamma_F$ and $d\Gamma^{V\ell\bar{\ell}'}/\Gamma_V$ indicate 
that they have to be evaluated in their respective frames. 
We assume that the production of the heavy fermion pair is through $QCD$ 
interactions. Since QCD conserves parity, the produced fermion $F$ is 
unpolarized. 
This allows the following simplification, after a partial integration over the 
$F\bar{F}$ phase space :
\begin{equation}\label{eq:17}
  d\sigma_{\lambda'\lambda}\rightarrow 
\frac{1}{f_{avg}^{F}}\frac{d\sigma_{F\bar{F}}}{d\bar{\Omega}}\delta_{
\lambda'\lambda}d\bar{\Omega}
\end{equation}
where $f_{avg}^{F}=2$ and $d\sigma_{F\bar{F}}/d\bar{\Omega}$ is the differential 
cross section for the production of $F\bar{F}$ pair, in the PCM frame. 
Substituting this in Eq.~\ref{eq:16}, 
we get,
\begin{equation}\label{eq:18}  
\hat{\sigma}=f_{avg}^{V}\times
\int\frac{d\sigma_{F\bar{F}}}{d\bar{\Omega}}\sum_{\{i\}}
R^{\dagger}_{i'k'}\left(\frac{d\Gamma^{FfV}_{\lambda\lambda';k'k}}{\Gamma_F}
\right)_{F}R_{ki}\left(\frac{d\Gamma^{V\ell\bar{\ell}'}_{ii'}}{\Gamma_V}
\right)_V d\bar{\Omega}.
\end{equation}
In the rest frame of $V$, the matrix $d\Gamma^{V\ell\bar{\ell}'}$  is given, 
after partial integration over phase space, by
\begin{equation}\label{eq:19} 
\frac{d\Gamma^{V\ell\bar{\ell}'}}{d\Omega^V_{\ell}}=\Gamma^{V\ell\bar{\ell}'}
\frac{\rho^{V}_{ii'}}{4\pi}
\end{equation}
where $d\Omega^V_{\ell}=\sin\theta_{\ell} d\theta_{\ell} d\phi_{\ell}$ and 
$\theta$, $\phi$ define the direction of motion of $\ell$ in the rest frame of 
$V$. The matrix $\rho^{V}$ is a matrix with $\sum_{i}\rho_{ii}^{V}=1$ and can be 
parameterized as follows:
\begin{equation}\label{eq:20}
  \rho^{V} =
  \begin{pmatrix}
    \frac{1+c^2_{\theta_{\ell}}+2\alpha c_{\theta_{\ell}}}{4} & 
\frac{s_{\theta_{\ell}}(\alpha+c_{\theta_{\ell}})}{2\sqrt{2}}e^{i\phi_{\ell}} & 
\frac{(1-c^2_{\theta_{\ell}})}{4}e^{2i\phi_{\ell}}\\

\frac{s_{\theta_{\ell}}(\alpha+c_{\theta_{\ell}})}{2\sqrt{2}}e^{-i\phi_{\ell}} & 
\frac{s^2_{\theta_{\ell}}}{2} & 
\frac{s_{\theta_{\ell}}(\alpha-c_{\theta_{\ell}})}{2\sqrt{2}}e^{i\phi_{\ell}}\\
    \frac{(1-c^2_{\theta_{\ell}})}{4}e^{-2i\phi_{\ell}} & 
\frac{s_{\theta_{\ell}}(\alpha-c_{\theta_{\ell}})}{2\sqrt{2}}e^{-i\phi_{\ell}} & 
\frac{1+c^2_{\theta_{\ell}}-2\alpha c_{\theta_{\ell}}}{4}
        \end{pmatrix}  
\end{equation}

where $s_x=\sin x$, $c_x=\cos x$. 
In the above equation, $\alpha=((g_L^{\ell})^2-(g_R^{\ell})^2)/((g_L^{\ell})^2+(g_R^{\ell})^2)$ with $g_L^{\ell}$, and  $g_R^{\ell}$ denoting
the left and right chiral couplings of the 
$V\ell\bar{\ell}'$ vertex.  

In the case where the polarization of $V$ is measured in the rest frame of $F$ 
obtained by a sequence of Lorentz transformations starting from the PCM frame,
\begin{equation}\label{eq:21}
  \mathrm{PCM}\; 
\mathrm{frame}\xrightarrow{h^{-1}(p_F^{PCM})}\mathrm{F}\;\mathrm{rest}\;\mathrm{
frame}\xrightarrow{h^{-1}(p_F^{F})}\mathrm{V}\;\mathrm{rest}\,\mathrm{frame},
\end{equation}
the helicity states of $V$ do not undergo helicity rotation. In this case,
\begin{equation}\label{eq:22} 
\left(\frac{d\Gamma^{FfV}}{d\Omega^F}\right)_{F}=\Gamma^{FfV}\frac{\rho^{F}_{i'i
}}{4\pi} 
\end{equation}
where $\Gamma^{FfV}$ is the partial width for the decay $F\rightarrow fV$, 
$\rho^{F}$ is a matrix with constant elements, and  $\sum_{i}\rho_{ii}^{F}=1$. 
In this case, the expression for $\hat{\sigma}$ becomes,
\begin{equation}\label{eq:23}
  \hat{\sigma}=\int \sigma_{F\bar{F}} \times\frac{f_{avg}^{V}}{4\pi}\sum_{\{i\}}\rho^{F}_{i'i}\rho^{V}_{ii'}
d\Omega^{V}_{\ell}
\times B.R(F\rightarrow V f)\times 
B.R(V\rightarrow 
\ell\bar{\ell}')
\end{equation}
where $B.R$ denotes a branching ratio. From the above expression, differential 
angular distribution of $\ell$ in the rest frame of $V$ can be obtained:
\begin{equation}\label{eq:24} 
\frac{1}{\sigma}\frac{d\sigma}{d\Omega^{V}_{\ell}}=\frac{f_{avg}^{V}}{4\pi}\sum_
{\{i\}}\rho^{F}_{i'i}\rho^{V}_{ii'}.
\end{equation}
This implies that the matrix $\rho^{F}$ can be regarded as the density matrix 
for the production of $V$ in the rest frame of $F$. The matrix $\rho^{F}$ can be 
parameterized as follows~\cite{Boudjema:2009fz}: 
\begin{equation*}
\rho^F=
 \begingroup
 \setlength\arraycolsep{0.1pt}
  \begin{pmatrix}
    \frac{1}{3}+\frac{P_z}{2}+\frac{T_{zz}}{\sqrt{6}} & 
\frac{P_x-iP_y}{2\sqrt{2}}+\frac{T_{xz}-iT_{yz}}{\sqrt{3}} & 
\frac{T_{xx}-T_{yy}-2iT_{xy}}{\sqrt{6}}\\
    \frac{P_x+iP_y}{2\sqrt{2}}+\frac{T_{xz}+iT_{yz}}{\sqrt{3}} & 
\frac{1}{3}-\frac{2T_{zz}}{\sqrt{6}} & 
\frac{P_x-iP_y}{2\sqrt{2}}-\frac{T_{xz}-iT_{yz}}{\sqrt{3}}\\
    \frac{T_{xx}-T_{yy}+2iT_{xy}}{\sqrt{6}} & 
\frac{P_x+iP_y}{2\sqrt{2}}-\frac{T_{xz}+iT_{yz}}{\sqrt{3}} &   
\frac{1}{3}-\frac{P_z}{2}+\frac{T_{zz}}{\sqrt{6}}
  \end{pmatrix}.
  \endgroup
\end{equation*}
In the above expression, $P_{i}$ and $T_{ij}$ ($i=x,y,z$) contain the 
information of the polarization of the vector boson $V$. 
$T_{ij}$ is a symmetric traceless tensor ($T_{xx}+T_{yy}+T_{zz}=0$) and $P_{i}$ 
is a vector. 

Substituting the expression for  
$\left(\frac{d\Gamma^{FfV}}{d\Omega^F}\right)_{F}$ and 
$\left(\frac{d\Gamma^{V\ell\bar{\ell}'}}{d\Omega^V_{\ell}}\right)_{V}$ in 
Eq.~\ref{eq:18}, we get,
\begin{align}\label{eq:26}
  \hat{\sigma} &= \frac{1}{4\pi}B.R(F\rightarrow 
fV)B.R(V\rightarrow\ell\bar{\ell}')\times\\\nonumber
&\frac{f_{avg}^{V}}{4\pi}\int 
\frac{d\hat{\sigma}_{F\bar{F}}}{d\bar{\Omega}} 
\operatorname{Tr}(R^{\dagger}\rho^{F}R\rho^{V})d\Omega 
d\Omega_{\ell}d\bar{\Omega} 
\end{align}
where $d\Omega=\sin\theta d\theta d\phi$, $d\Omega_{\ell}=\sin\theta_{\ell} 
d\theta_{\ell} d\phi_{\ell}$, and 
$d\bar{\Omega}=\sin\bar{\theta} d\bar{\theta} 
d\bar{\phi}$.
Evaluating the trace in Eq.~\ref{eq:26}, we get, after substituting the 
expressions for the matrices $\rho^{V}$ and $\rho^{F}$,
\begin{align}\label{eq:27}
  \frac{d\hat{\sigma}}{d\Omega_{\ell}}=&B.R.(F\rightarrow 
fV)B.R(V\rightarrow\ell\bar{\ell}')\times \frac{f_{avg}^{V}}{8\pi}\times \\\nonumber
  &\Bigg[\int\left(\frac{2}{3}-\frac{T_{zz}}{\sqrt{6}}
\left(c^2_{\omega}-\frac{1}{2}s^2_{\omega}\right)\right)\Big(\frac{d\hat{\sigma}
_{F\bar{F}}}{d\bar{\Omega}}\Big)\frac{d\Omega}{4\pi}d\bar{\Omega}\\\nonumber      
&+\alpha c_{\theta_{\ell}}\int\Big(\frac{d\hat{\sigma}_{F\bar{F}}}{d\bar{\Omega}}
\Big)(c_{\omega} P_z)\frac{d\Omega}{4\pi}d\bar{\Omega} \\\nonumber
&+c^2_{\theta_{\ell}}\sqrt{\frac{3}{2}}\int\Big(\frac{d\hat{\sigma}_{F\bar{F}}}{
d\bar{\Omega}}\Big)\Big(T_{zz}\left(c^2_{\omega}-\frac{1}{2}s^2_{\omega}
\right)\Big)\frac{d\Omega}{4\pi}d\bar{\Omega}\\\nonumber               
&+\alpha s_{\theta_{\ell}}c_{\phi_{\ell}}\Big(\int\Big(\frac{d\hat{\sigma}_{F\bar
{F}}}{d\bar{\Omega}}\Big)(s_{\omega} c_{\chi} 
P_z)\frac{d\Omega}{4\pi}d\bar{\Omega} \Big)\\\nonumber          
&+\alpha s_{\theta_{\ell}}s_{\phi_{\ell}}\Big(\int\Big(\frac{d\hat{\sigma}_{F\bar
{F}}}{d\bar{\Omega}}\Big)(-s_{\omega} s_{\chi} 
P_z)\frac{d\Omega}{4\pi}d\bar{\Omega}\Big) \\\nonumber           
&+s_{\theta_{\ell}}c_{\theta_{\ell}}c_{\phi_{\ell}}\Big(\int\Big(\frac{d\hat{
\sigma}_{F\bar{F}}}{d\bar{\Omega}}\Big)(\sqrt{6}s_{\omega} c_{\omega} c_{\chi} 
T_{zz})\frac{d\Omega}{4\pi}d\bar{\Omega} \Big)\\\nonumber
&+s_{\theta_{\ell}}c_{\theta_{\ell}}s_{\phi_{\ell}}\Big(\int\Big(\frac{d\hat{
\sigma}_{F\bar{F}}}{d\bar{\Omega}}\Big)(-\sqrt{6}s_{\omega} c_{\omega} c_{\chi} 
T_{zz})\frac{d\Omega}{4\pi}d\bar{\Omega}\Big) \\\nonumber                 
&+s^2_{\theta_{\ell}}s_{2\phi_{\ell}}\Big(\int\Big(\frac{d\hat{\sigma}_{F\bar{
F}}}{d\bar{\Omega}}\Big)(-\frac{1}{4}\sqrt{\frac{3}{2}}s^2_{\omega}s_{2\chi} 
T_{zz})\frac{d\Omega}{4\pi}d\bar{\Omega}\Big)\\\nonumber                   
&+s^2_{\theta_{\ell}}c_{2\phi_{\ell}}\Big(\int\Big(\frac{d\hat{\sigma}_{F\bar{
F}}}{d\bar{\Omega}}\Big)(\frac{1}{4}\sqrt{\frac{3}{2}}s^2_{\omega} c_{2\chi} 
T_{zz})\frac{d\Omega}{4\pi}d\bar{\Omega}\Big)\Bigg]
\end{align}
where $c_{x}=\cos x$, $s_{x}=\sin x$. The angle $\omega$ is given by the 
following expressions:
\begin{align}\label{eq:28}
\cos{\omega}&=\frac{(\beta+\bar{\beta}\cos\theta)}{\sqrt{\beta^2+\bar{\beta}
^2-\beta^2\bar{\beta}^2\sin^2\theta+2\beta\bar{\beta}\cos\theta}},\\\nonumber
\sin{\omega}&=\frac{\bar{\beta}\sin\theta}{\gamma\sqrt{\beta^2+\bar{\beta}
^2-\beta^2\bar{\beta}^2\sin^2\theta+2\beta\bar{\beta}\cos\theta}}.
\end{align}
Note that the angle $\omega$ is independent of the direction of motion of $F$ in 
the PCM frame, i.e, independent of $\bar{\theta}$ and $\bar{\phi}$.
We have the following expressions for $\chi$:
\begin{align}\label{eq:29}
\cos\chi&=\cos\phi\cos\Delta\phi-\sin\Delta\phi\cos\bar{\theta}\sin\phi,
\\\nonumber
\sin\chi&=-\frac{\sin\Delta\phi\sin\bar{\theta}\sqrt{\bar{\beta}^{2}
+\beta^2-\bar{\beta}^{2}\beta^2\sin^2\theta+2\bar{\beta}\beta\cos\theta}}{\sqrt{
1-\bar{\beta}^2}\beta\sin\theta},
\end{align}
where $\Delta\phi=\bar{\phi}-\phi^{PCM}$, $\phi^{PCM}$ is a function of 
$\beta,\bar{\beta},\theta,\phi$ and $\bar{\theta}$. The terms with $\cos\chi$ 
and/or $\sin\chi$ drop out
after integration over the azimuthal angle $\phi_{\ell}$ of the lepton, since 
these terms always appear with factors such as $\sin\phi_{\ell}$, $\cos 
2\phi_{\ell}$, etc. Equation~\ref{eq:27}
becomes,
\begin{align}\label{eq:30}
\frac{d\hat{\sigma}}{d\cos\theta_{\ell}}&= 2\pi\times B.R.(F\rightarrow 
fV)B.R(V\rightarrow\ell\bar{\ell}')\times\frac{f_{avg}^{V}}{8\pi}\times\\\nonumber
& \Bigg[\int\left(\frac{2}{3}-\frac{T_{zz}}{\sqrt{6}}
\left(c^2_{\omega}-\frac{1}{2}s^2_{\omega}\right)\right)\Big(\frac{d\hat{\sigma}
_{F\bar{F}}}{d\bar{\Omega}}\Big)\frac{d\Omega}{4\pi}d\bar{\Omega}\\\nonumber
&+\alpha c_{\theta_{\ell}}\int\Big(\frac{d\hat{\sigma}_{F\bar{F}}}{d\bar{\Omega}}
\Big)(c_{\omega} P_z)\frac{d\Omega}{4\pi}d\bar{\Omega} \\\nonumber
&+c^2_{\theta_{\ell}}\sqrt{\frac{3}{2}}\int\Big(\frac{d\hat{\sigma}_{F\bar{F}}}{
d\bar{\Omega}}\Big)\Big(T_{zz}\left(c^2_{\omega}-\frac{1}{2}s^2_{\omega}
\right)\Big)\frac{d\Omega}{4\pi}d\bar{\Omega}\Bigg].
\end{align}

The terms with $\cos\omega$ and $\sin\omega$ are independent of the direction of 
motion of $F$ ($\bar{\theta}$, $\bar{\phi}$) in the PCM frame. This means that 
the 
integration $\int (d\hat{\sigma}_{F\bar{F}}/d\bar{\Omega}) d\bar{\Omega}$ can be 
performed independently to give a factor $\sigma_{F\bar{F}}$. The angle $\omega$ 
is also independent of the azimuthal
angle ($\phi$) of the vector boson $V$ in the rest frame of $F$. Hence, a 
partial integration of $d\Omega=d\cos\theta d\phi$ can be performed 
independently. The simplified Eq.~\ref{eq:30}
reads,
\begin{align}\label{eq:31}
\frac{1}{\hat{\sigma}_{F\bar{F}}}\frac{d\hat{\sigma}}{d\cos\theta_{\ell}}&=  
B.R.(F\rightarrow fV)B.R(V\rightarrow\ell\bar{\ell}')\frac{f_{avg}^{V}}{4}\\\nonumber
&\Bigg[\int\left(\frac{2}{3}-\frac{T_{zz}}{\sqrt{6}}
\left(c^2_{\omega}-\frac{1}{2}s^2_{\omega}\right)\right)\frac{1}{2}
d\cos\theta\\\nonumber
&+\alpha c_{\theta_{\ell}}\int(c_{\omega} P_z)\frac{1}{2}d\cos\theta \\\nonumber
&+c^2_{\theta_{\ell}}\sqrt{\frac{3}{2}}\int\Big(T_{zz}\left(c^2_{\omega}-\frac{1
}{2}s^2_{\omega}\right)\Big)\frac{1}{2}d\cos\theta\Bigg].
\end{align}
Defining
 \begin{align}\label{eq:32}
  \mathcal{T}_{zz}(\bar{\beta})&=\left(\frac{1}{2}\int 
d\cos\theta\Big(\cos^2\omega 
-\frac{1}{2}\sin^2\omega\Big)\right)T_{zz},\\\nonumber
 &=\frac{1}{4}\left(\int d\cos\theta (3\cos^2\omega -1)\right) T_{zz}\\\nonumber
  \mathcal{P}_z(\bar{\beta})&=\left(\frac{1}{2}\int d\cos\theta 
\cos\omega\right)P_z,\\\nonumber
\end{align}
we rewrite the above equation as
\begin{align}\label{eq:32a}
 \frac{1}{\hat{\sigma}_{F\bar{F}}}\frac{d\hat{\sigma}}{d\cos\theta_{\ell}}&= 
B.R.(F\rightarrow fV)B.R(V\rightarrow\ell\bar{\ell}')\\\nonumber
 &\times  
\frac{f_{avg}^{V}}{4}\Bigg[\frac{2}{3}-\frac{\mathcal{T}_{zz}(\bar{\beta})}{
\sqrt{6}}+\alpha\cos\theta_{\ell}\mathcal{P}_z(\bar{\beta})\\\nonumber
 &+\sqrt{\frac{3}{2}}\cos^2\theta_{\ell}\mathcal{T}_{zz}(\bar{\beta})\Bigg].
\end{align}
Comparing this expression with the expression for the azimuthal-averaged angular 
distribution of a decay product $\ell$ of a vector boson in its rest 
frame~\cite{Rahaman:2016pqj},
 \begin{equation}\label{eq:33}    
\frac{1}{\sigma}\frac{d\sigma}{d\cos\theta_{\ell}}=\frac{f_{avg}^{V}}{4}\Big[
(\frac{2}{3}-\frac{T_{zz}}{\sqrt{6}})+\alpha 
P_z\cos\theta_{\ell}+\sqrt{\frac{3}{2}}T_{zz}\cos^2\theta_{\ell}\Big],
  \end{equation}
  where $\sigma$ is the cross section for the production of $V$, 
we interpret the quantities $\mathcal{P}_z(\bar{\beta})$ and 
$\mathcal{T}_{zz}(\bar{\beta})$ as the polarization parameters of $V$, for a 
given parton level event,
as seen by a direct boost from the parton center of mass frame to the rest frame 
of $V$. Note that such a simplification does not arise had the factors involving 
$\chi$
been kept in Eq.~\ref{eq:30}, since they are functions of all the angular 
variables in the problem and hence the integration would have become a 
multi-dimensional one. In 
such a case, this method does not offer any advantage over a Monte Carlo 
simulation to extract the polarization parameters of $V$. 
 
 Before we complete the derivation of the expression of polarization estimators 
(see Sec.~\ref{sec:5}), we discuss the expressions in  Eq.~\ref{eq:32}. 
 Performing the integrations over $\cos\theta$, we get,
 \begin{widetext}
 \begin{eqnarray}\label{eq:34} 
\mathcal{P}_{z}(\bar{\beta})&=&\frac{1}{2\beta^2\bar{\beta}}\left[2\bar{\beta}
-(1-\beta^2)\log\left(\frac{1+\bar{\beta}}{1-\bar{\beta}}\right)\right]P_z\,\,\,
\,(\bar{\beta}<\beta)\nonumber\\
&=&\frac{1}{2\beta^2\bar{\beta}}\left[2\beta-(1-\beta^2)\log\left(\frac{1+\beta}
{1-\beta}\right)\right]P_z\,\,\,(\bar{\beta}>\beta),\\\nonumber
&=&\frac{1}{\beta^3}\left[\beta-(1-\beta^2)\frac{1}{2}\log\left(\frac{1+\beta}{
1-\beta}\right)\right]P_z\,\,\,\,(\bar{\beta}=\beta),
 \end{eqnarray}
 \end{widetext}
where $\bar{\beta}$ is the velocity of the parent particle in the PCM frame and 
$\beta$ is the velocity of the vector boson $V$ in the rest frame of the parent 
particle. Similarly, we get 
the expression for $\mathcal{T}_{zz}(\bar{\beta})$:
\begin{widetext}
\begin{align}\label{eq:35}
\mathcal{T}_{zz}(\bar{\beta})&=\frac{T_{zz}}{8\beta^3\bar{\beta}}\Bigg[-4\beta\bar{
\beta}(-3+\beta^2)\\\nonumber
&+3\log\left(\frac{1-\beta\bar{\beta}-\sqrt{1-\beta^2}\sqrt{1-\bar{\beta}^2}}{
1+\beta\bar{\beta}-\sqrt{1-\beta^2}\sqrt{1-\bar{\beta}^2}}
\right)\left(\left(2-\sqrt{\frac{1-\beta^2}{1-\bar{\beta}^2}}
\right)(1-\beta^2)-\sqrt{1-\beta^2}\sqrt{1-\bar{\beta}^2}\right)\\\nonumber
&-3\log\left(\frac{1-\beta\bar{\beta}+\sqrt{1-\beta^2}\sqrt{1-\bar{\beta}^2}}{
1+\beta\bar{\beta}+\sqrt{1-\beta^2}\sqrt{1-\bar{\beta}^2}}
\right)\left(\left(2+\sqrt{\frac{1-\beta^2}{1-\bar{\beta}^2}}
\right)(1-\beta^2)+\sqrt{1-\beta^2}\sqrt{1-\bar{\beta}^2}\right)\Bigg], 
(\bar{\beta} \neq \beta)\\\nonumber
& = 
-\frac{T_{zz}}{2\beta^4}\Big[\beta^2(-3+\beta^2)-3(1-\beta^2)\log(1-\beta^2)\Big
], (\bar{\beta}=\beta).
 \end{align}
\end{widetext}

\subsection{Discussion}\label{ssec:2a} 
\begin{figure}
\includegraphics[scale=0.47,keepaspectratio=true]{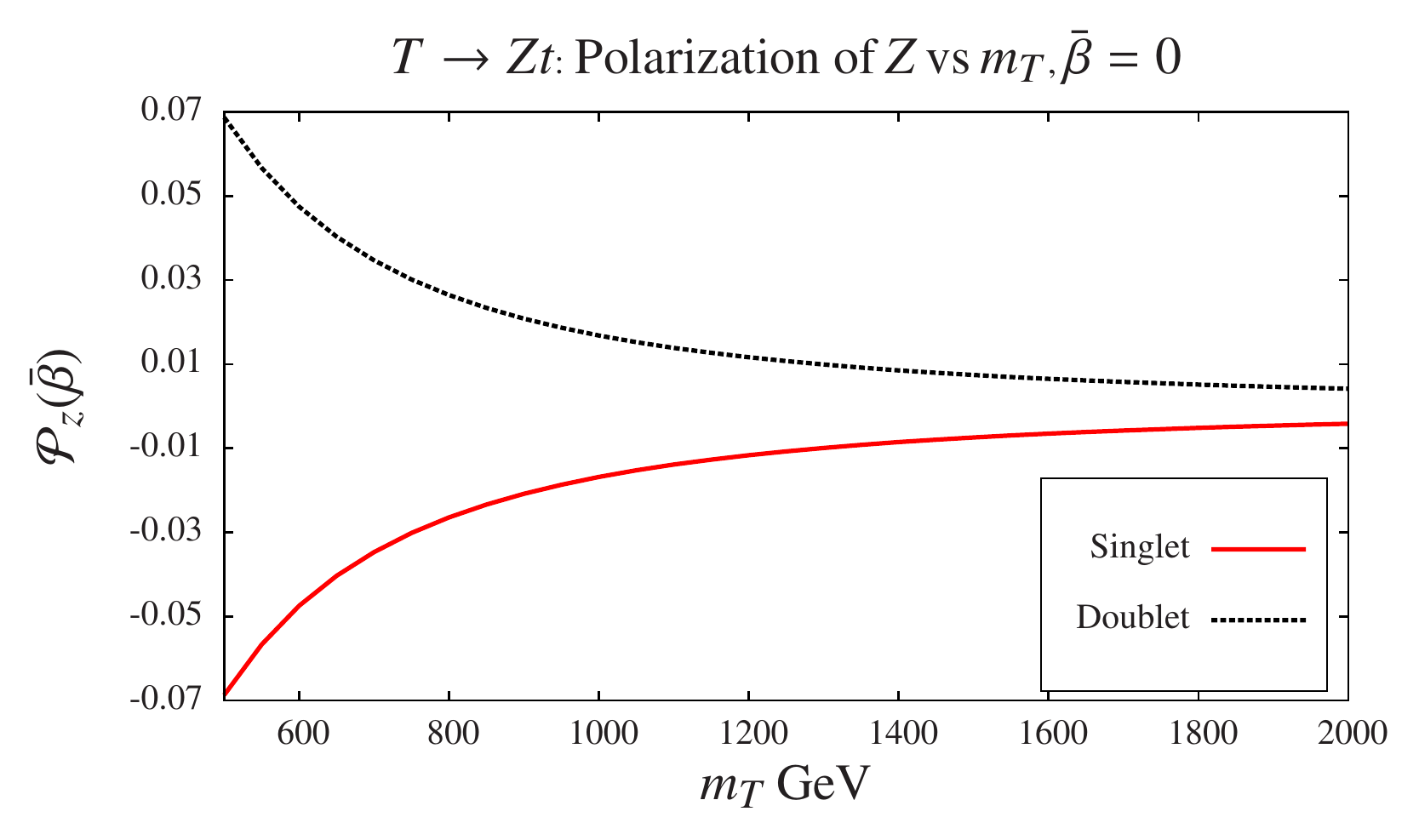}
\includegraphics[scale=0.47,keepaspectratio=true]{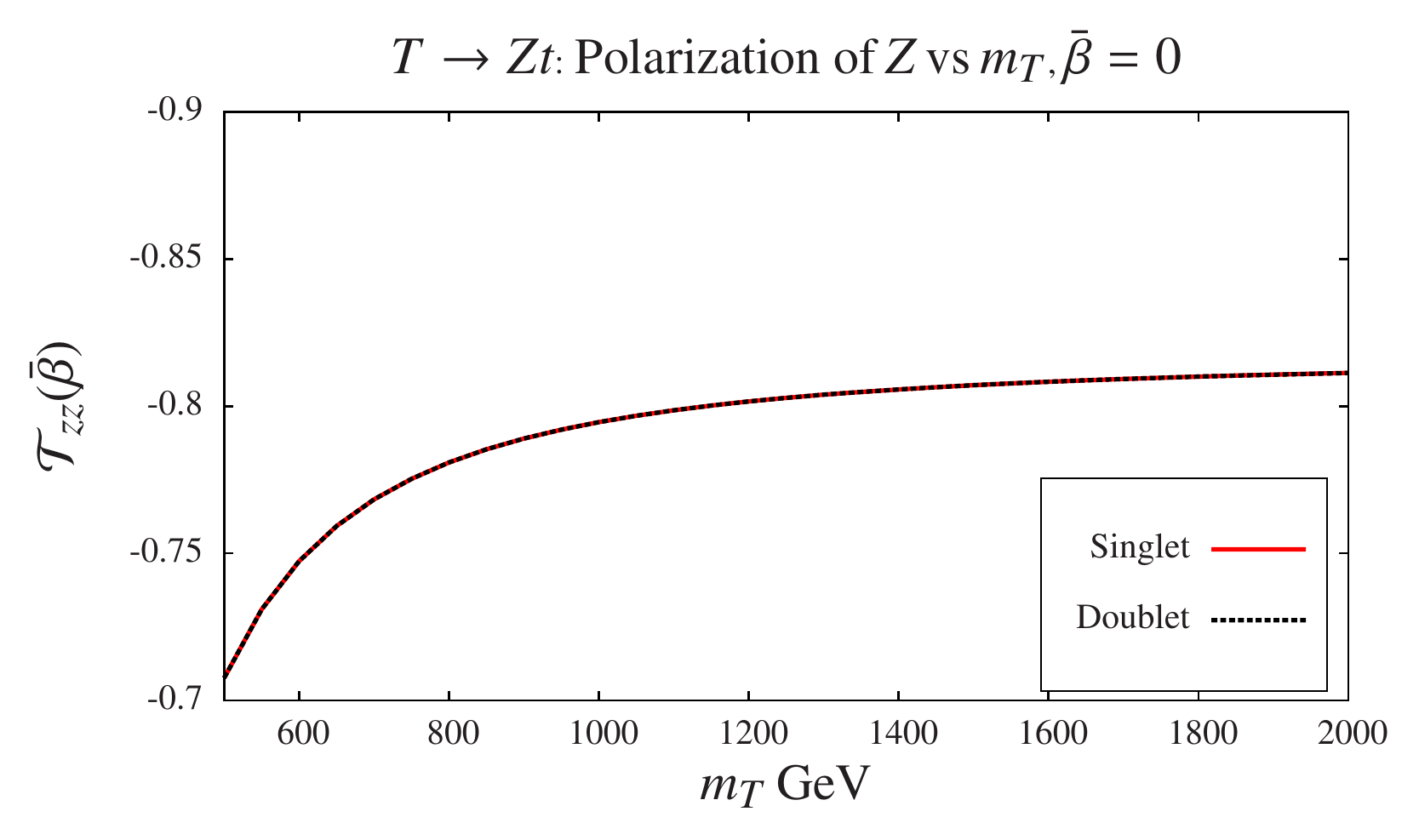}
\caption{The value of polarization parameters of $Z$, in the rest frame of $T$, 
as a function of $m_T$. The solid (dashed) lines correspond to the singlet (doublet) model. The 
remaining parameters of the two models  are taken as in Table.~\ref{tab:1}. In the figure on the right panel, the lines
corresponding to the two models are identical and appear merged.} \label{fig:1}
\end{figure}

 We study the expressions Eq.~\ref{eq:34} and Eq.~\ref{eq:35} for models of 
vector-like quarks given in Sec.~\ref{sec:3} where such decays are possible. We also consider the case of top 
decays in the SM. 
One can evaluate the value of $\mathcal{P}_z(\bar{\beta})$ and $\mathcal{T}_{zz}(\bar{\beta})$ at any 
given value of the velocity of the heavy fermion $\bar{\beta}$ using 
Eq.~\ref{eq:34} and Eq.~\ref{eq:35}
and the expression for $\beta$:
\begin{equation}\label{eq:42}
 \beta = \frac{K(1,\xi_V,\xi_f)^{1/2}}{1+\xi_V-\xi_f},
\end{equation}
obtained in the rest frame of $F$. 
  Figure~\ref{fig:1} shows the polarization parameters $P_z$ and $T_{zz}$ in the 
rest frame of $T$ ($\bar{\beta}=0$) as a function of $m_T$, for the two models  
given in Table~\ref{tab:1}.  One can see that, for large values of $m_T$, the 
value of $P_z$ tends to zero while that of $T_{zz}$ tends to a constant value. 
This can be understood
from Eq.~\ref{eq:41} (with $\xi_V\rightarrow \xi_Z$, $\xi_f\rightarrow \xi_t$). 
In the limit of large $m_T$, $|P_z|\rightarrow 2\xi_Z/(1-\xi_t)\rightarrow 0$ 
and 
$T_{zz}\rightarrow -\sqrt{2/3}\approx -0.82$, due to the fixed masses of $Z$ and 
$t$. Note that both $T_{zz}$ and $P_z$ are independent of the value of the 
mixing angles since
one of the two couplings of $ZtT$, $g_L$ and $g_R$, is always zero for the two 
models considered in this work. This leads to results of
$\mathcal{T}_{zz}(\bar{\beta})$ in the singlet model being identical to that of the doublet model, for any given value of $\bar{\beta}$ and $m_T$,
which can be seen on the right panel of Fig.~\ref{fig:1}. 

\begin{figure}
\includegraphics[scale=0.47,keepaspectratio=true]{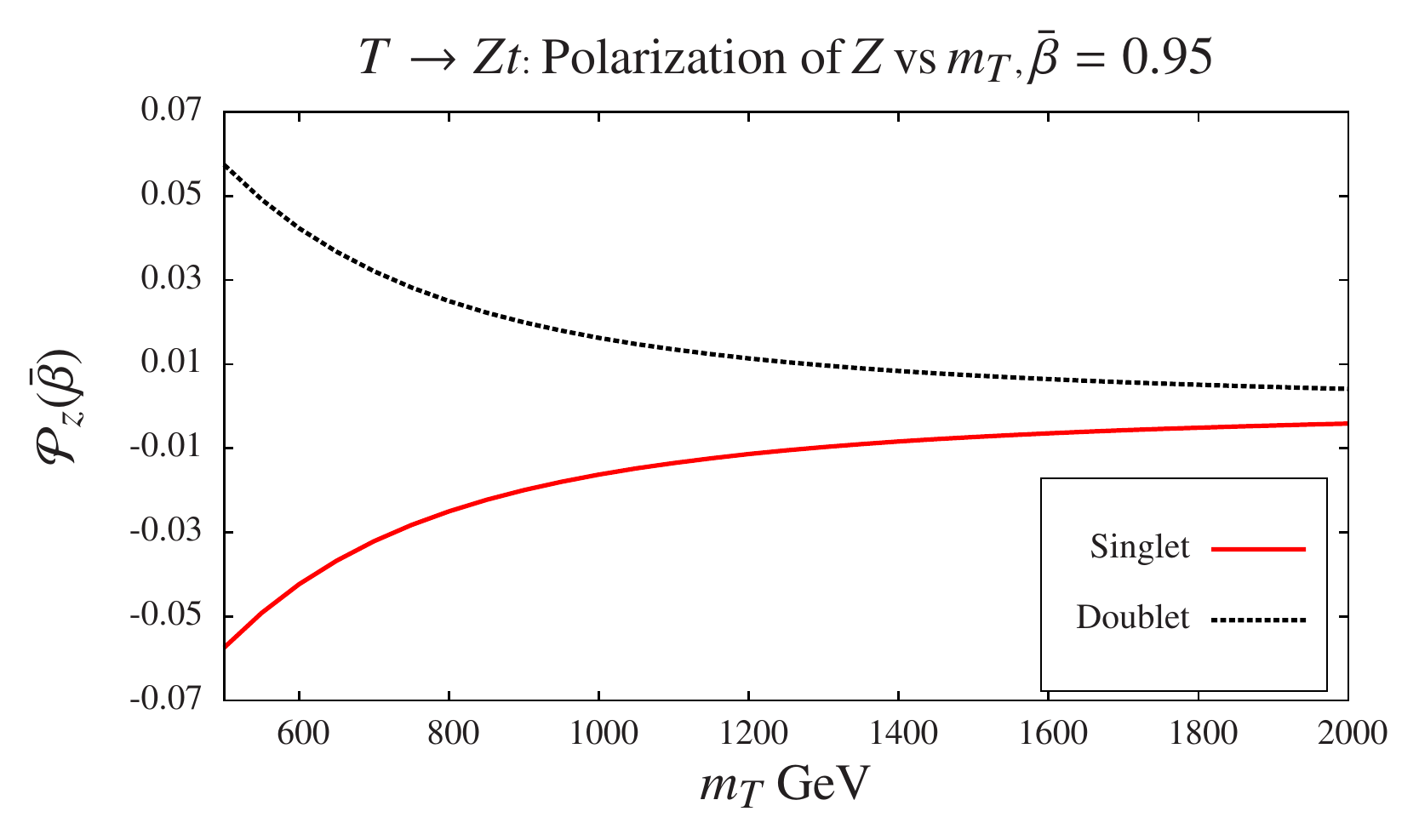}
\includegraphics[scale=0.47,keepaspectratio=true]{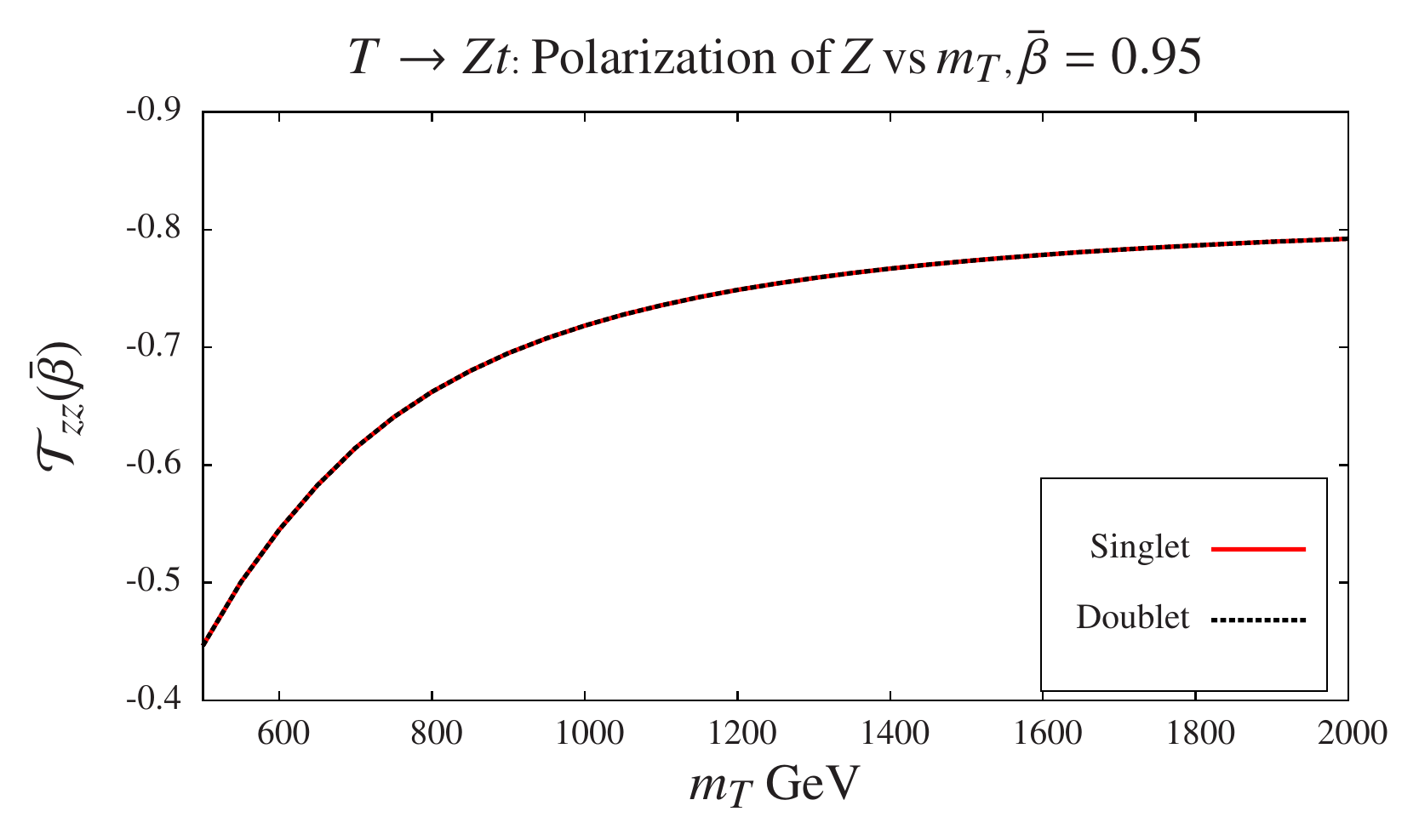}
\caption{The value of polarization parameters of $Z$ in a frame where $T$ moves with a velocity
$\bar{\beta}=0.95$ as a function of $m_T$. The solid (dashed) lines correspond to the singlet (doublet) model. The 
remaining parameters of the two models  are taken as in Table.~\ref{tab:1}.
In the figure on the right panel, the lines
corresponding to the two models are identical and appear merged.}\label{fig:2}
\end{figure}

  Figure~\ref{fig:2} shows the values of polarization parameters 
$P_z(\bar{\beta})$ and $T_{zz}(\bar{\beta})$ for $\bar{\beta}=0.95$, for the two 
models. Comparing Fig.~\ref{fig:2} with 
 Fig.~\ref{fig:1}, one  can see that the value of $T_{zz}$ significantly differs 
in the two cases, for each model, when the mass of the heavy fermion $T$ is low. 
Such a large difference
 is not observed in the case of $P_z$. In other words, the effect of the boost 
of the rest frame of $T$ relative to the PCM frame is more important in $T_{zz}$ 
than $P_z$.  However, such 
 effects decrease when the mass of $T$ becomes larger since the velocity 
$\beta$ of $Z$ in the rest frame of $T$ approaches unity (for a fixed 
$\bar{\beta}$). In the limit $\beta\rightarrow 1$, Eq.~\ref{eq:34} and 
Eq.~\ref{eq:35}
 give $\mathcal{P}_z(\bar{\beta})\rightarrow P_{z}$ and 
 $\mathcal{T}_{zz}(\bar{\beta})\rightarrow T_{zz}$, for a fixed $\bar{\beta}$. In other words, the values of $\mathcal{P}_z(\bar{\beta})$ and $\mathcal{T}_{zz}(\bar{\beta})$ tend to remain close to their values in the rest frame of $F$, for
 any value of $\bar{\beta} < \beta$ , thus reducing their sensitivity to $\bar{\beta}$ ($< \beta $).  
  
\begin{figure}
\includegraphics[scale=0.47,keepaspectratio=true]{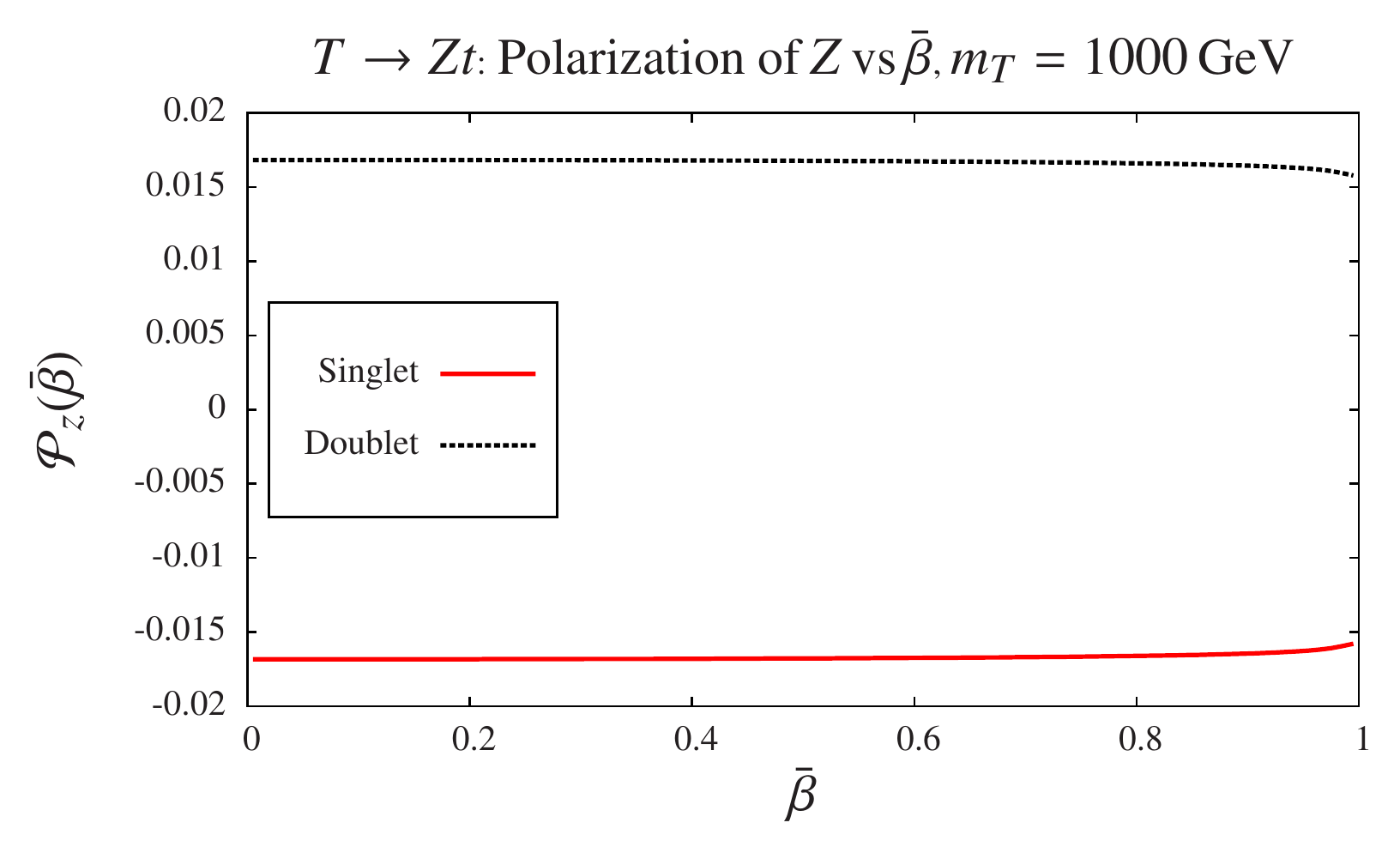}
\includegraphics[scale=0.47,keepaspectratio=true]{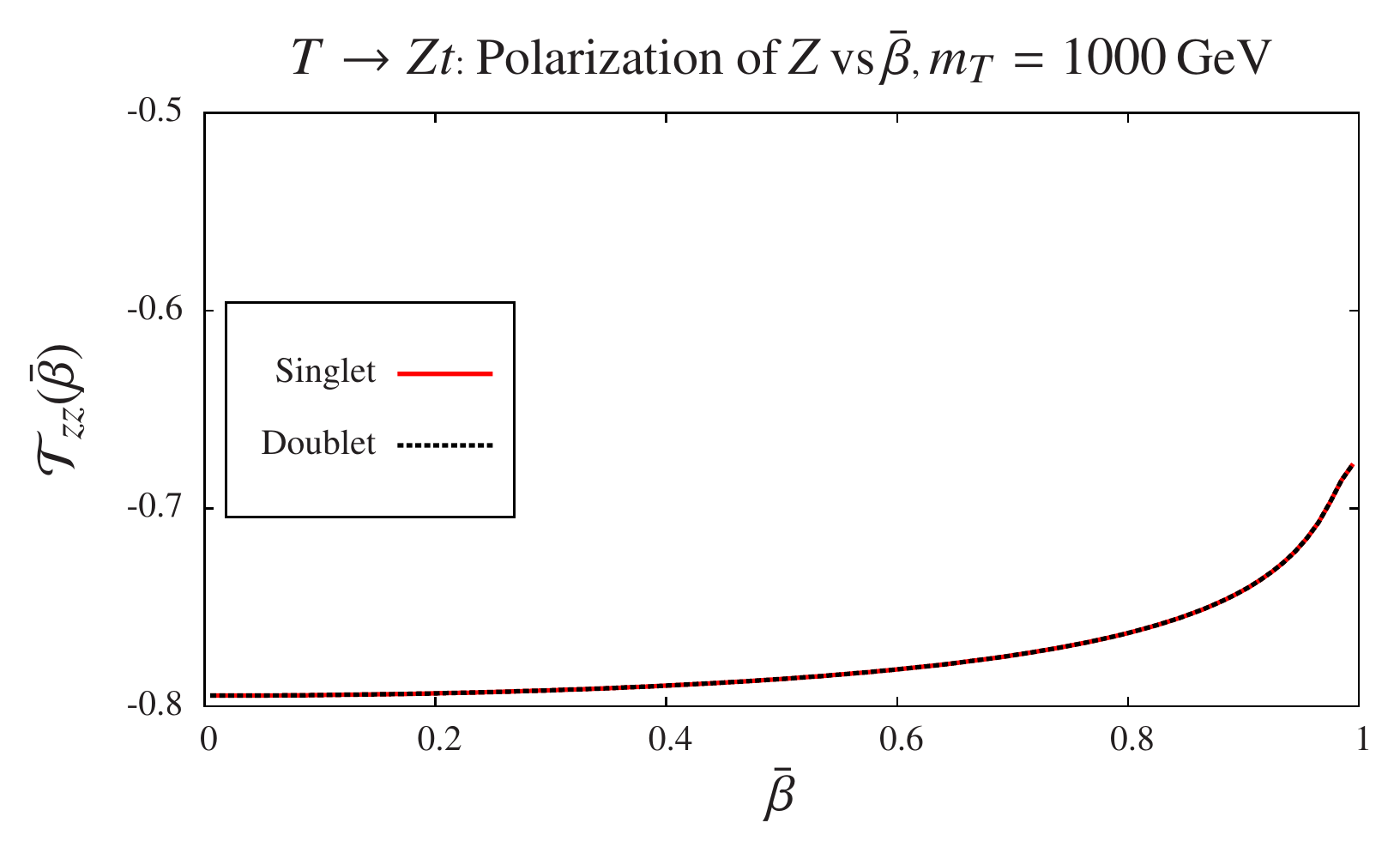}
\caption{The value of polarization parameters of $Z$ as a function of  velocity $\bar{\beta}$ of the vector-like quark $T$,
 with a mass $m_T=1000$ GeV. The solid (dashed) lines correspond to the singlet (doublet) model. The 
remaining parameters of the two models  are taken as in Table.~\ref{tab:1}.
In the figure on the right panel, the lines
corresponding to the two models are identical and appear merged.
}\label{fig:3}
\end{figure}

     The dependence of the polarization parameters $\mathcal{P}_z(\bar{\beta})$ and 
$\mathcal{T}_{zz}(\bar{\beta})$ is shown in Fig.~\ref{fig:3}, for the two models and for 
$m_T=1000$ GeV. One can see that the effect of the boost of the rest frame of $T$ 
relative to the PCM frame ($\bar{\beta}$) is stronger at large values of  
$\bar{\beta}$, as it should be.       

\begin{figure}
\includegraphics[scale=0.5,keepaspectratio=true]{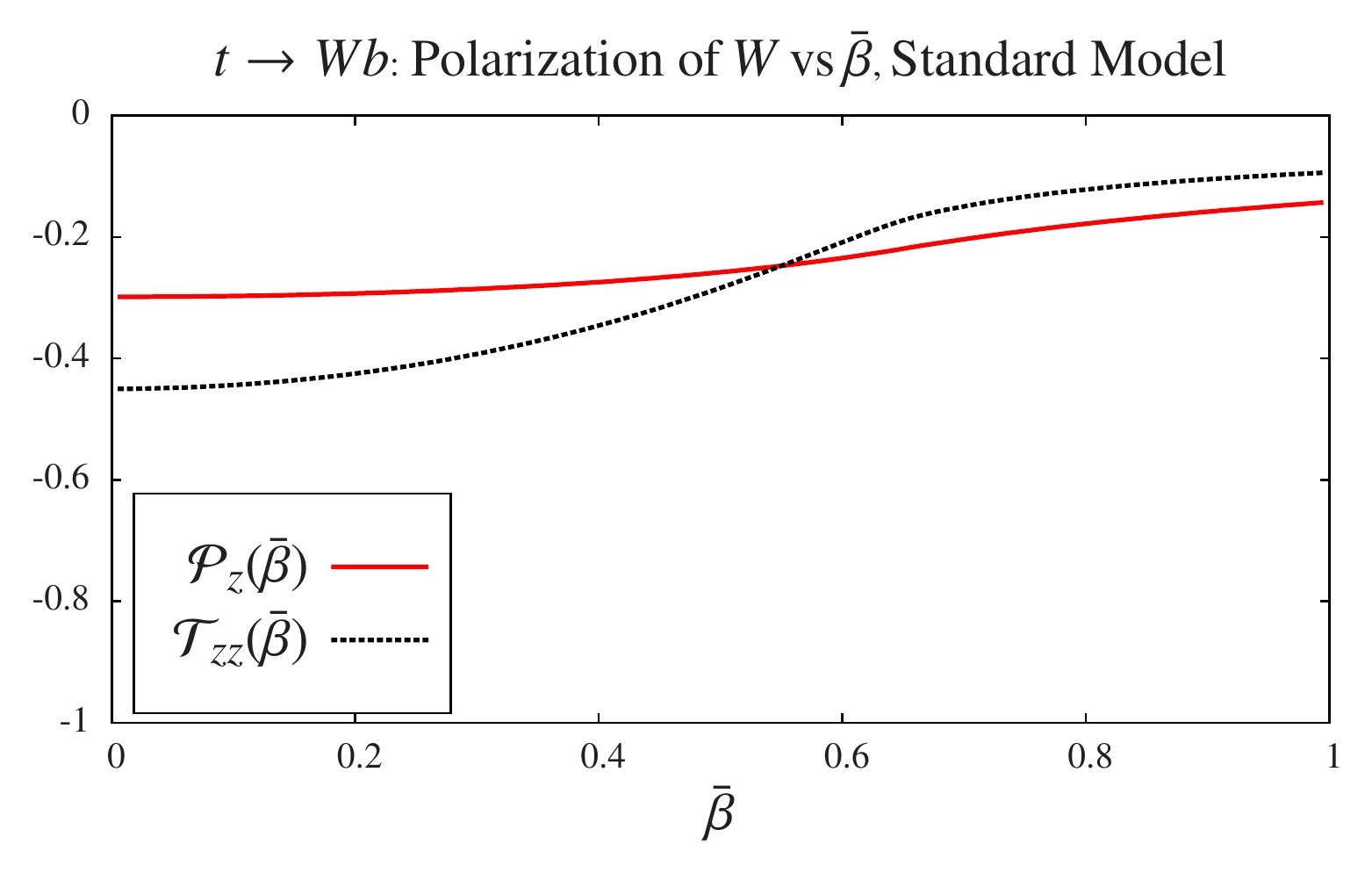}
\caption{The polarization parameters of $W$ in the decay of top quark in the SM as a function of velocity of top quark $\bar{\beta}$.
The solid (dashed) lines correspond to $\mathcal{P}_z(\bar{\beta})$ ($\mathcal{T}_{zz}(\bar{\beta})$).}\label{fig:4}
\end{figure}

For completeness, we show in Fig~\ref{fig:4}, the polarization parameters of 
$W$in the  top decay, in the SM, as a function of $\bar{\beta}$. In this case, 
due  to the relatively lighter parent particle, the top, the dependence of $\mathcal{P}_z(\bar{\beta})$ 
and $\mathcal{T}_{zz}(\bar{\beta})$ on $\bar{\beta}$ (the velocity of the top in the PCM frame) are
stronger even for moderate values of $\bar{\beta}$. 

\section{Polarization estimators}\label{sec:5}
To obtain expressions for polarization parameters of $V$ at the level of 
$pp$ collisions, the expressions in Eq.~\ref{eq:34} and Eq.~\ref{eq:35} on $\mathcal{P}_z(\bar{\beta})$ and $\mathcal{T}_{zz}(\bar{\beta})$
need to be convoluted over the parton distribution functions (pdfs). 
This is required since the parton distribution functions and the cross section 
determine the $\bar{\beta}$ distribution in the PCM frame. Defining the cross 
section for the process $pp\rightarrow F\bar{F}\rightarrow 
\bar{F}fV\rightarrow\bar{F}f\ell\ell'$ by
\begin{equation}\label{eq:34a}
  \sigma=\int dx_1dx_2 f_{p1/p}(x_1)f_{p2/p}(x_2)\hat{\sigma}
\end{equation}
where $\hat{\sigma}=\hat{\sigma}_{F\bar{F}}B.R(F\rightarrow fV)B.R(V\rightarrow 
\ell\ell')$, we get the expressions for the polarization parameters as
\begin{align}\label{eq:35a}
  \mathcal{P}_z^{NW}&=\frac{1}{\sigma}\int dx_1dx_2 f_{p1/p}(x_1)f_{p2/p}(x_2) 
\mathcal{P}_z(\bar{\beta}),\\\nonumber
  \mathcal{T}_{zz}^{NW}&=\frac{1}{\sigma}\int dx_1dx_2 
f_{p1/p}(x_1)f_{p2/p}(x_2) \mathcal{T}_{zz}(\bar{\beta}).
 \end{align}
The superscript NW refers to the fact that we have used the NWA for $F$ (for $V$ 
it is applicable). This expression can also be written as an average over the 
$\bar{\beta}$ distribution
as it is the only variable in the problem (the c.m. energy $\sqrt{\hat{s}}$ can 
be traded for $\bar{\beta}$ through $\bar{\beta}=\sqrt{1-4m_F^2/\hat{s}}$).
\begin{align}\label{eq:36a}
 \mathcal{P}_z^{NW}&=\int \frac{1}{\sigma}\frac{d\sigma}{d\bar{\beta}} 
\mathcal{P}_z(\bar{\beta}) d\bar{\beta},\\\nonumber
 \mathcal{T}_{zz}^{NW}&= \int \frac{1}{\sigma}\frac{d\sigma}{d\bar{\beta}} 
\mathcal{T}_{zz}(\bar{\beta}) d\bar{\beta}
\end{align}
where $1/\sigma d\sigma/d\bar{\beta}$ is the normalized velocity distribution of 
$F$ in the PCM frame. 

   To covert the above expressions into the corresponding lab frame quantities, 
we replace the $\bar{\beta}$ distribution with the normalized velocity 
distribution of $F$ in the 
   lab frame $1/\sigma d\sigma /d\beta^{lab}$ and the polarization parameters by 
$P_z(\beta^{lab})$ and $T_{zz}(\beta^{lab})$ with $\beta^{lab}$ being the 
velocity of $F$ in
   the lab frame. In this case, the rest frame of $F$ is understood to be 
obtained by a direct boost of $\beta^{lab}$ from the lab frame. Hence, the final 
expressions for the 
   polarization estimators of $V$ produced in the decay of $F$,  in the lab 
frame are
\begin{align}\label{eq:37}
 \mathcal{P}_z^{NW}&=\int \frac{1}{\sigma}\frac{d\sigma}{d\beta^{lab}} 
\mathcal{P}_z(\beta^{lab}) d\beta^{lab},\\\nonumber
 \mathcal{T}_{zz}^{NW}&= \int \frac{1}{\sigma}\frac{d\sigma}{d\beta^{lab}} 
\mathcal{T}_{zz}(\beta^{lab}) d\beta^{lab}.
\end{align}
  This equation can be interpreted as the average of  
$\mathcal{P}_z(\beta^{lab})$ and $\mathcal{T}_{zz}(\beta^{lab})$ over all the 
events with a weighting factor $(1/\sigma) d\sigma/d\beta^{lab}$.
We now present the numerical validation the expressions for the polarization
estimators of $V$ given Eq.~(\ref{eq:37}). We  generate
events for the process $pp\rightarrow T\bar{T}$ followed by the decay of $T$ 
into $Z$ and $t$
with $Z$ further decaying to $\ell\bar{\ell}$, using {\tt 
MadGraph}~\cite{Alwall:2014hca}. The events correspond to the two models in 
Table~\ref{tab:1} 
For the models of $T$ concerned, due to the strong constraints on the couplings, 
the width of $T$ remains much smaller compared to its mass
throughout the range of mass i.e. 900 GeV to 2000 GeV. This means that the NWA 
is a good approximation throughout the mass range considered and the expression 
in Eq.~\ref{eq:34} and Eq.~\ref{eq:35} can be expected to be valid.   We also 
considered the case of $W$ polarization in top decays both in the case of 
the singlet and the doublet Models and  in the case of the SM \footnote{The results for $W$ polarization in the case of the SM and in the case of the singlet model 
are not shown as they are identical to 
the other two cases. This is due to the fact that the $tbW$ couplings ($g_L$ and $g_R$) in the two vector-like quark models are very close to 
the corresponding SM values as a result of strong constraints on the $t-T$ and $b-B$ mixing angles.}. For this purpose, 
we generated events for $pp\rightarrow t\bar{t}$ and allowed the top to decay to 
$W$ and $b$ with $W$ 
further decaying to $\bar{\ell}\nu$. This provides an additional verification of 
our method. 
We use the value of velocity ($\beta^{lab}$) of the heavy fermion ($T$ or $t$)
in the lab frame, from the generated sample and compute the quantities 
$\mathcal{P}_z(\beta^{lab})$ and $\mathcal{T}_{zz}(\beta^{lab})$ for each event and 
obtain an average over the entire event sample. This is equivalent to the use of 
Eq.~\ref{eq:37}.  This method yields the values of $\mathcal{P}_z^{NW}$ and $\mathcal{T}_{zz}^{NW}$, since the heavy 
fermion is assumed to be on-shell due to the NWA.
   
\begin{figure}
\centering
\includegraphics[scale=0.47,keepaspectratio=true]{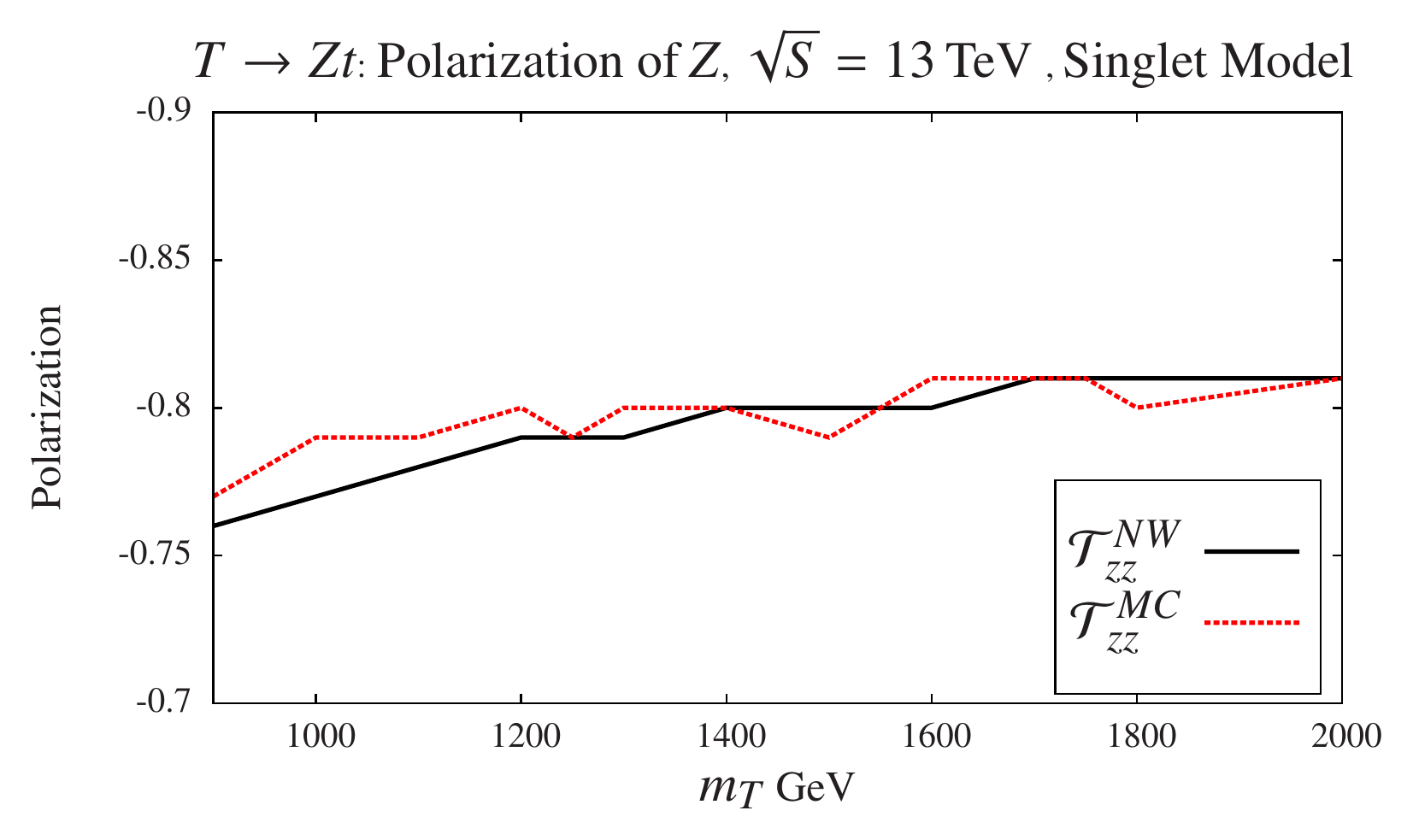}
\includegraphics[scale=0.47,keepaspectratio=true]{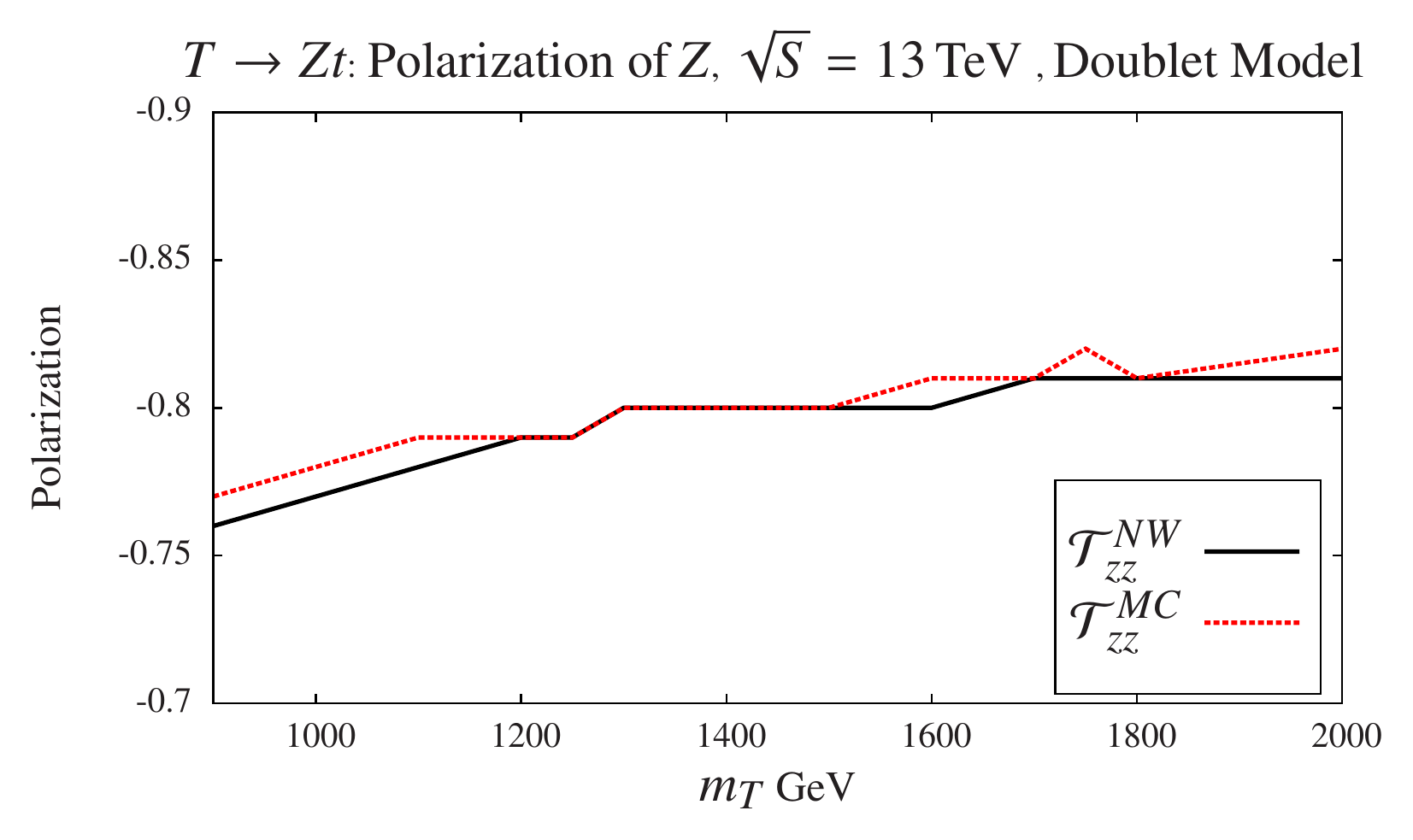}
\caption{A comparison of the two polarization estimators $\mathcal{T}_{zz}^{NW}$(solid) and $\mathcal{T}_{zz}^{MC}$(dashed) of $Z$ produced in the decay $T\rightarrow Z t$
for different choices of mass of $T$ ($m_T$). The left (right) panel corresponds to the singlet (doublet) model. 
The $pp$ collision center of mass energy $\sqrt{S}$ is taken as 13 TeV. The remaining parameters for the models are taken as in Table.~\ref{tab:1}.}\label{fig:num1}
\end{figure}

\begin{figure}
\centering
\includegraphics[scale=0.47,keepaspectratio=true]{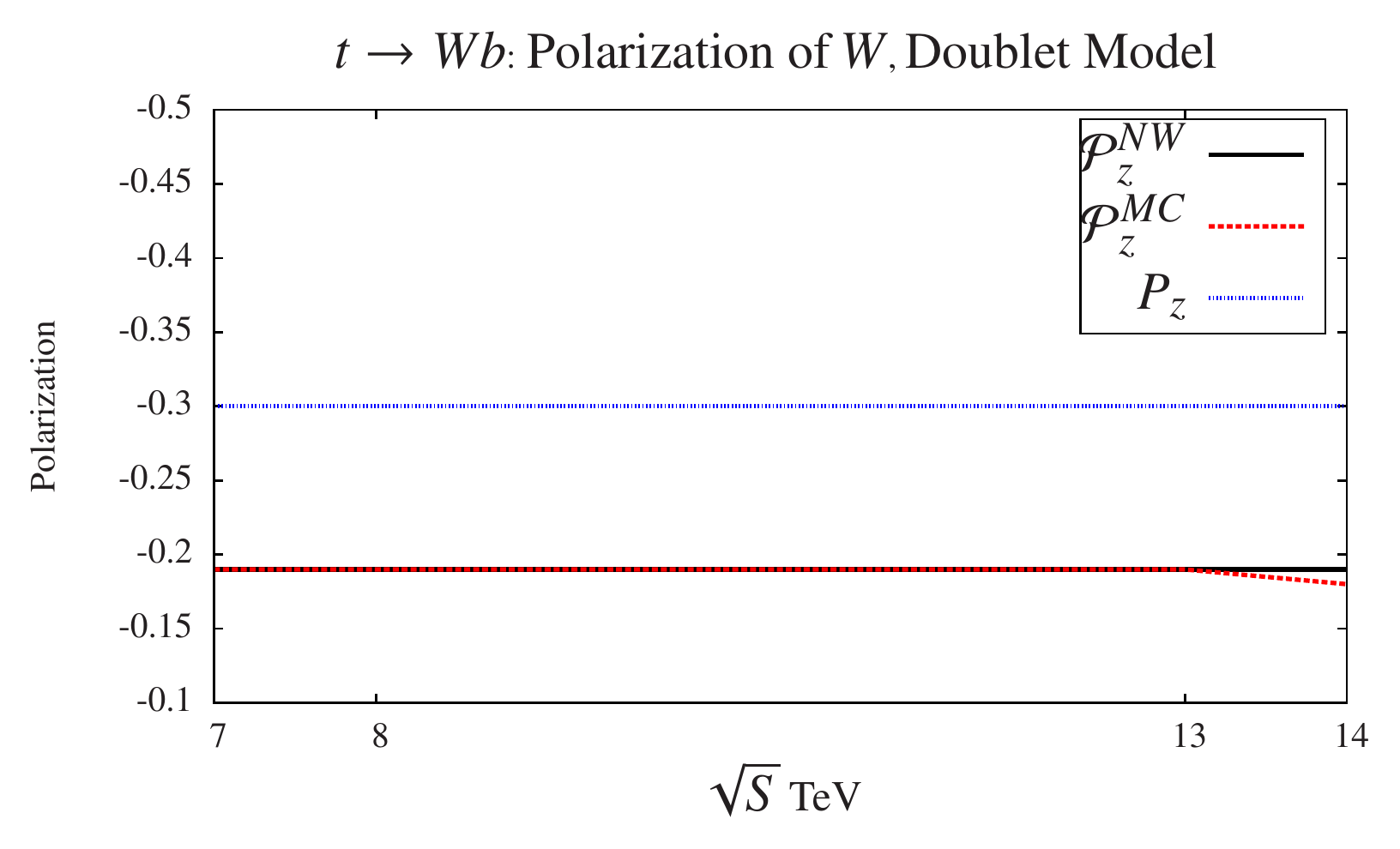}
\includegraphics[scale=0.47,keepaspectratio=true]{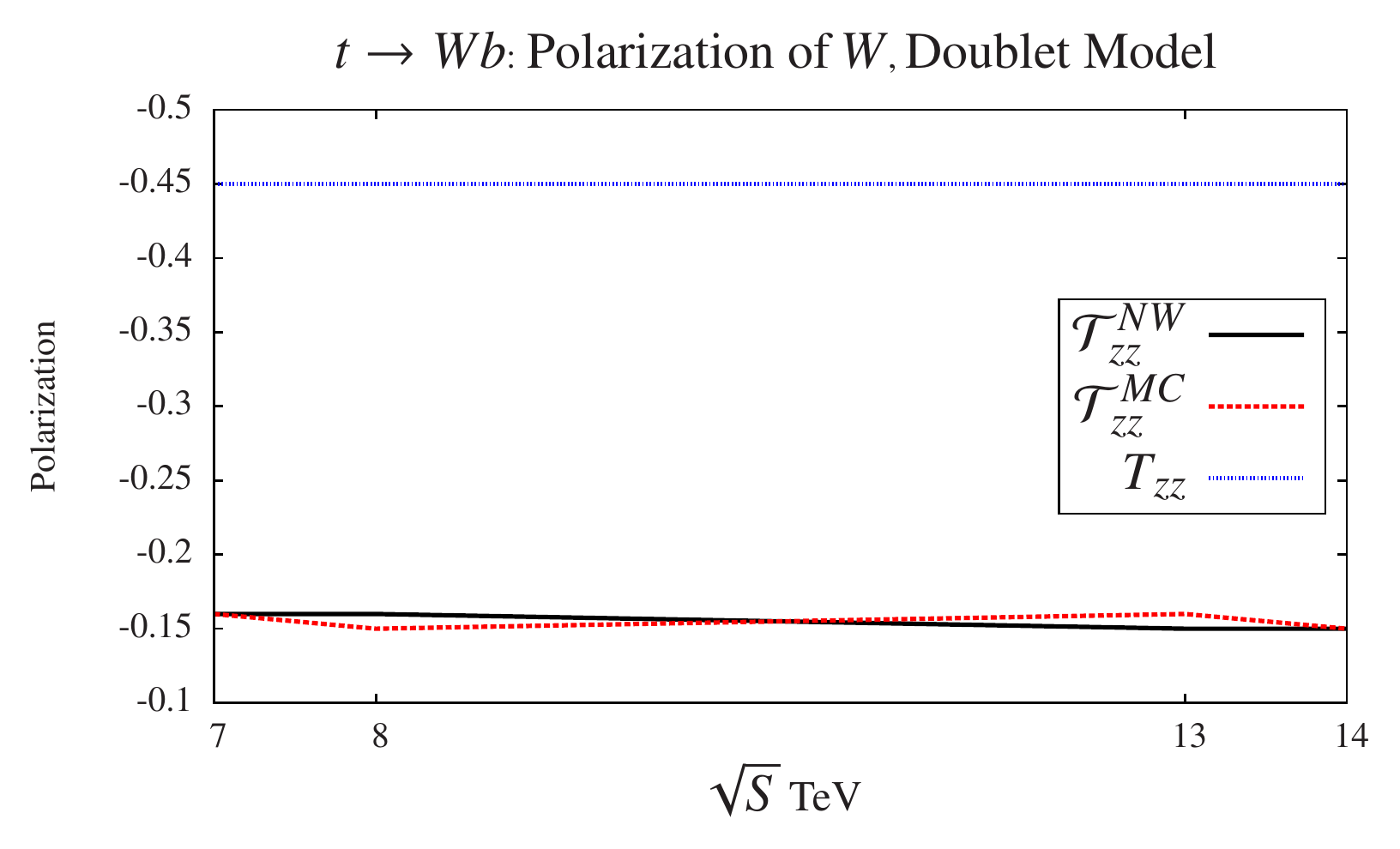}
\caption{A comparison of the polarization estimators of $W$ produced in the decay $t\rightarrow W b$
 in the doublet model of vector-like quark $T$, for different choices of $pp$ collision center of mass energy $\sqrt{S}$.  The left (right) panel 
 compares $\mathcal{P}_{z}^{NW}$ (solid) and $\mathcal{P}_{z}^{MC}$(dashed) ($\mathcal{T}_{zz}^{NW}$ (solid) and $\mathcal{T}_{zz}^{MC}$ (dashed)) along with the $W$ polarization 
 in the top quark rest frame $P_z$ (dotted) ($T_{zz}$ (dotted)).   
The mass of $T$ ($m_T$) is taken as 900 GeV.  The remaining parameters for the models are taken as in Table.~\ref{tab:1}.}\label{fig:num2}
\end{figure}

 The value of polarization parameters can be directly extracted from the 
Monte Carlo event samples through the use of angular asymmetries of the lepton 
$\ell$
 from the $Z$ decay~\cite{Rahaman:2016pqj}. Consider the asymmetries $A_z$ and 
$A_{zz}$ defined by
 \begin{equation}\label{eq:38}  
A_z=\frac{\sigma(\cos\theta_{\ell}>0)-\sigma(\cos\theta_{\ell}<0)}{
\sigma(\cos\theta_{\ell}>0)+\sigma(\cos\theta_{\ell}<0)}
 \end{equation}
and 
\begin{equation}\label{eq:39}
 A_{zz}=\frac{\sigma(\sin 3\theta_{\ell}>0)-\sigma(\sin 
3\theta_{\ell}<0)}{\sigma(\sin 3\theta_{\ell}>0)+\sigma(\sin 3\theta_{\ell}<0)}
\end{equation}
where $\theta_{\ell}$ is the polar angle of the lepton (from the $Z$ decay) in 
the rest frame of $Z$ that is obtained by a direct boost from the lab frame. 
$\sigma$ is the cross section for the production of $V$ followed by 
$V\rightarrow \ell\bar{\ell}'$ and is the same as $\sigma$ in Eq.~\ref{eq:34a}. 
These asymmetries can be directly related to the values of
polarization parameters of $Z$ (see Eq.~\ref{eq:33}):
performing the convolutions with parton distribution functions,
we get
\begin{align}\label{eq:40a}
 A_z&=\frac{3\alpha }{4}\mathcal{P}_z^{MC},\\\nonumber
 T_{zz}&=\frac{3}{8}\sqrt{\frac{3}{2}}\mathcal{T}_{zz}^{MC}
\end{align}
where the $MC$ superscript refers to the fact the polarization parameters are 
extracted from a Monte Carlo simulation. Note that in the case of $W$ produced 
in the top decay,
the lepton $\ell$ corresponds to a neutrino ($W\rightarrow \bar{\ell}\nu$), 
which is unobservable. Hence, one need to use the asymmetries of $\bar{\ell}$ in 
place of the above mentioned
asymmetries of $\ell$. Since the neutrino and the anti-charged lepton 
$\bar{\ell}$ have equal and  opposite momenta , 
$\theta_{\bar{\ell}}=\pi-\theta_{\ell}$. This means that there
is an additional factor of ($-1$) in the first expression of Eq.~\ref{eq:40a}.
   Figure~\ref{fig:num1} compares the value of polarization parameter 
$\mathcal{T}_{zz}^{NW}$  described above 
and the direct extraction
   from Monte Carlo simulation $\mathcal{T}_{zz}^{MC}$ for the decay 
$T\rightarrow Zt$. One can see that these values agree to within a few percent.    
The value of the other polarization parameters $\mathcal{P}_z^{NW}$, $\mathcal{P}_z^{MC}$ are not shown  as 
their 
   values are  close to zero
   in both the models, as shown in Fig.~\ref{fig:3}. Their numerical values in both the models are provided in Table~\ref{tab:2}. 
   As an additional proof that our method is 
valid, we consider the decay of top quark 
   in the two models. The results are shown in Fig.~\ref{fig:num2}, for the case 
of the doublet Model, for four choices of $pp$ center of mass energy $\sqrt{S}$. 
   One can see the excellent agreement between the two methods.  

\begin{table}
\begin{tabular}{>{\centering}p{3cm}>{\centering}p{3cm}>{\centering}p{3cm}>{\centering}p{3cm}>{\centering}p{3cm}}
\hline
\hline
 & 
\multicolumn{2}{c}{Singlet} & \multicolumn{2}{c}{Doublet}\tabularnewline
\cline{2-5}
$m_T$ (GeV) & $\mathcal{P}_z^{NW}$ & $\mathcal{P}_z^{MC}$ & $\mathcal{P}_z^{NW}$ & $\mathcal{P}_z^{MC}$ \tabularnewline
 \hline
 900 & -0.02 & -0.02 & 0.02 & 0.01\tabularnewline
1000 & -0.02 & -0.02 & 0.02 & 0.02 \tabularnewline
1250 & -0.01 & 0.01 & 0.01 & 0.02 \tabularnewline
1500 & -0.01 & -0.01 & 0.01 & 0.00 \tabularnewline
1750 & -0.01 & -0.01 & 0.01 & 0.01 \tabularnewline
2000 &  0.00 & 0.00 & 0.00 & -0.01 \tabularnewline
\hline
\hline
\end{tabular}
\caption{ A comparison of polarization estimators $\mathcal{P}_z^{NW}$ and $\mathcal{P}_z^{MC}$ of $Z$ produced in the decay $T\rightarrow Zt$, in the singlet and the doublet models,
for different values of the mass ($m_T$) of the vector-like quark $T$. The remaining parameters of the two models are taken as in Table.~\ref{tab:1}. The $pp$ center of 
mass energy $\sqrt{S}$ is taken as 13 TeV.}
\label{tab:2}
\end{table}

We have so far assumed  that the width of the mother particle is small compared 
to its mass. This justified
our application of the Narrow Width Approximation by which we have taken the 
on-shell mass of the mother particle
as its mass. Due to the strong constraints on the couplings of the vector-like 
quarks, in the two models that are
considered here the width of $T$ remains small ($\Gamma/m < 0.02$) compared to 
its mass throughout the mass range considered. 
This can be expected since the decays of $T$ are of electroweak type. However, 
we indicate how to extend the applicability
of our method to the case where the width of $T$ is large ($\Gamma/m\sim 0.1$). 
In this case,  non-resonant contributions to the process in question can not be neglected, in general. For example, the vector boson may 
come from a $t$-channel heavy fermion exchange 
rather than coming from the decay of the heavy fermion.
The kinematics of such a 
process is different from the decay process we are interested 
in. Moreover, there are additional spin-correlations between the production and 
the decay of the heavy fermion $T$ when the fermion
is off-shell~\cite{Vega:1995cc,Ballestrero:1994jn,Richardson:2001df}. These off-shell effects are not present in the narrow width case, since the application of NWA results in on-shell vector-like quark $F$. Hence, the 
extension of our method to the cases where $T$ has a finite width is, in 
general, highly non-trivial.
However, in the case where the non-resonant production and the additional 
spin-correlation effects are small, we can construct 
appropriately modified polarization estimators. Since, in this case, the only 
additional effect is the smearing of mass of the parent
particle, we take the invariant mass given by the four-momentum carried by its 
propagator as its mass. This mass can then be used in 
expressions such as Eq.~\ref{eq:34}, Eq.~\ref{eq:35} and the resulting 
polarization estimators are denoted as $\mathcal{P}_z^{BW}$ and 
$\mathcal{T}_{zz}^{BW}$ where the subscript $BW$ refers to the Breit-Wigner 
shape of the propagator used in Monte Carlo simulations.

\begin{figure}
\centering
\includegraphics[scale=0.45,keepaspectratio=true]{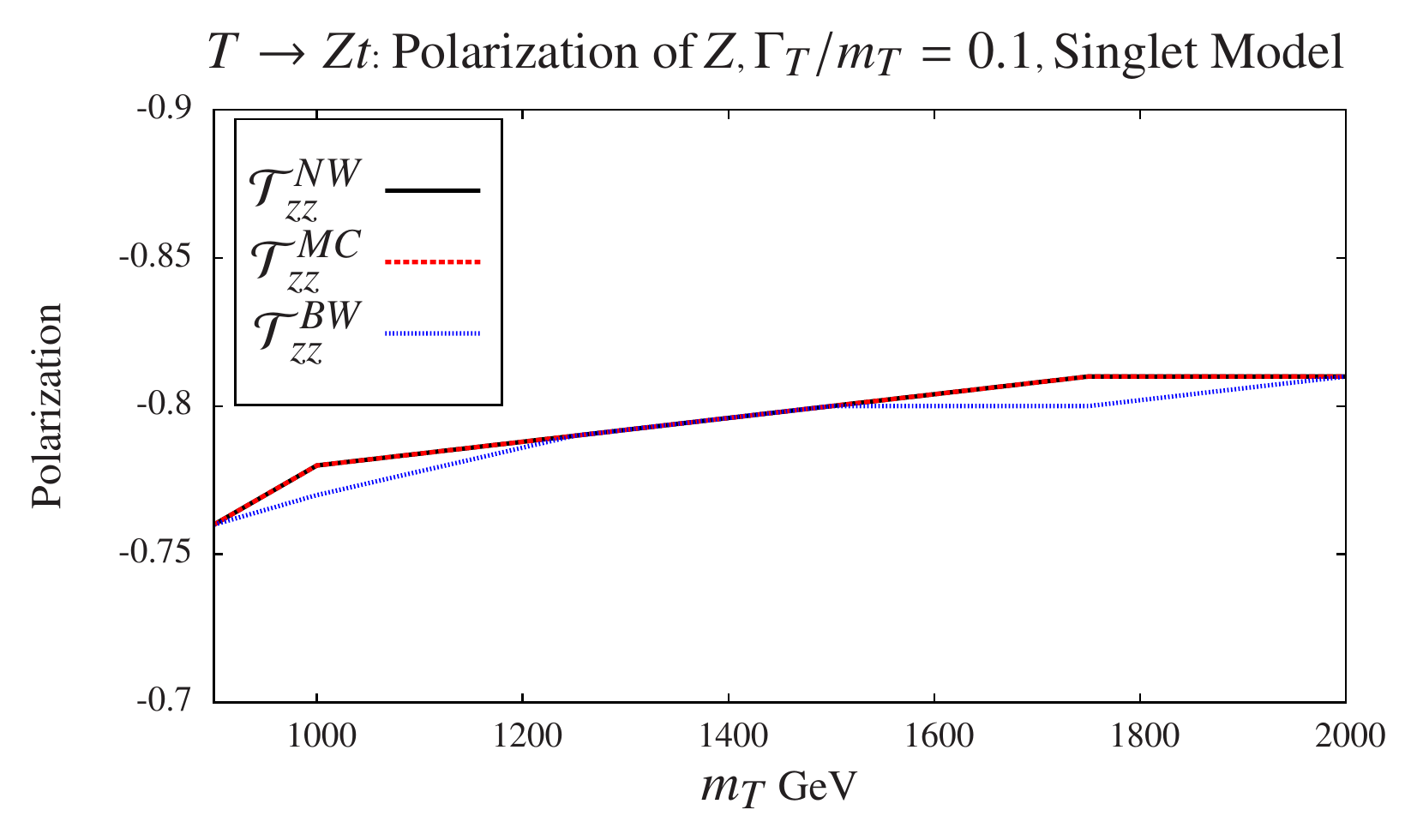}
\includegraphics[scale=0.45,keepaspectratio=true]{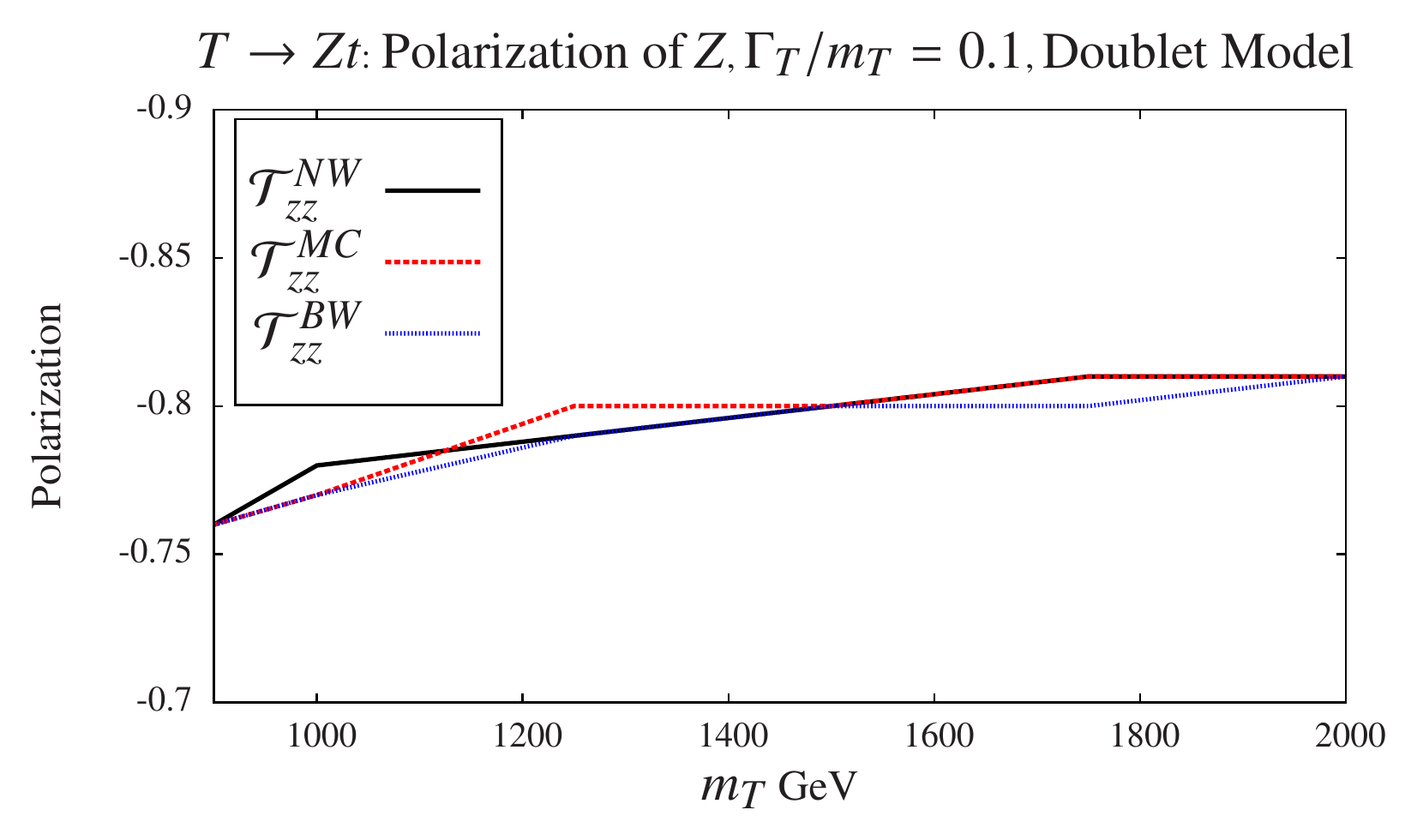}
\caption{A comparison of the polarization estimators $\mathcal{T}_{zz}^{NW}$ (solid), $\mathcal{T}_{zz}^{MC}$ (dashed) and $\mathcal{T}_{zz}^{BW}$ (dotted) of $Z$ produced in the decay $T\rightarrow Z t$
for different choices of mass of $T$ ($m_T$). The left (right) panel corresponds to the singlet (doublet) model. The width-to-mass ratio $\Gamma_T/m_T$ of the vector-like quark $T$
is taken to be 0.1. The $pp$ collision center of mass energy $\sqrt{S}$ is taken as 13 TeV. The remaining parameters for the models are taken as in Table.~\ref{tab:1}.}\label{fig:num3}
\end{figure}

A comparison of the two sets of estimators
is shown in Fig.~\ref{fig:num3}, for the two models given in Table~\ref{tab:1}  with the width-to-mass ratio of $10\%$ ($\Gamma_T/m_T=0.1$) for the vector-like quark $T$.
The corresponding plot for the other set of polarization estimators $\mathcal{P}_z^{NW}$, $\mathcal{P}_z^{MC}$ and $\mathcal{P}_z^{BW}$ are not shown as their values are
close to zero. One can see from Fig.~\ref{fig:num3} that both the estimators  and $\mathcal{T}_{zz}^{BW}$ agree with each other and with $\mathcal{T}_{zz}^{MC}$ to within a 
few percent. This shows that the estimator derived in the narrow width case can also be used when the parent particle has a finite width. This is justified  provided we neglect any off-shell spin and non-resonant contributions.

\section{Summary}\label{sec:6}
  In this work, we obtained expressions for the polarization parameters of a 
vector boson $V$ (such as $W,Z$) produced in the decay
  of a heavy fermion such as a possible vector-like quark or the top quark, both 
in the rest frame of the mother particle and in a frame
  where the mother particle is moving. Based on these expressions we constructed 
simple estimators of the polarization parameters
  requiring only the velocity distribution of the mother particle, apart from 
the necessary couplings and masses involved in the
  decay of the mother particle. Since the vector boson has both a vector and a 
tensor polarization, we construct two non-trivial estimators 
  one for each type of polarization, which survive in the azimuthal-averaged 
decay distribution of the vector boson. The estimators $\mathcal{P}^{NW}$, 
  $\mathcal{T}^{NW}_{zz}$ are derived under assumption that the width of the 
mother fermion is small compared its mass, applying the Narrow Width
  Approximation for the mother particle. The advantage of this method is the 
possibility of a quick estimate of polarization parameters of the vector boson,
  in frames such as the lab frame. The polarization estimated by this method  
can be measured experimentally without any requirement to reconstruct 
  any intermediate frame. We demonstrate the validity of this method in a set of 
models with a vector-like like quark  and also in the case
of the top decay in the SM.

We believe that this work 
  will aid the study of the vector-like quark phenomenology. This is because the 
polarization information carried by the vector boson from the decays of 
vector-like quarks 
  can be a probe of the coupling structure of the decay vertex. In addition to 
the case of vector-like quarks of narrow width, we also consider the cases where
  the vector-like quark has a finite width (width-to-mass ratio is taken to be 0.1) . We restrict ourselves to the cases 
where the non-resonant production of the same final states and additional spin
  correlation between the production and decay of the vector-like quark can be 
ignored. In this scenario, we propose two estimators $\mathcal{P}^{BW}_z$, 
  $\mathcal{T}^{BW}_{zz}$ similar to the previous estimators by introducing an 
additional convolution over the Breit-Wigner shape of the vector-like quark 
invariant
mass distribution. We validate the modified estimators and compare them with
the original estimators. We find that both the set of estimators provide equally
good approximations to the polarization parameters of $V$, in the finite width case.
\bibliographystyle{apsrev}
\bibliography{refs,ref2,refs-new,ref-diphoton}

\end{document}